\documentclass[twocolumn,showpacs,preprintnumbers,amsmath,amssymb,pra]{revtex4}
\usepackage[dvips]{graphics,graphicx,epstopdf}
\usepackage{mathrsfs}
\usepackage{graphics,graphicx,epstopdf}
\usepackage{pdfpages}
\usepackage{float}
\usepackage{hyperref}
\usepackage[hyphenbreaks]{breakurl}
\usepackage{braket}
\usepackage{natbib}
\begin{document}
\title{K-matrix formulation of
two-particle scattering in a wave guide in the presence of one-dimensional
spin-orbit coupling}
\author{Su-Ju Wang}
\email{sjwang@ou.edu}
\author{Q. Guan}
\email{gqz0001@gmail.com}
\author{D. Blume}
\email{doerte.blume-1@ou.edu}
\affiliation{Homer L. Dodge 
Department of Physics and Astronomy, The University of Oklahoma, 
440 West Brooks Street,
Norman, Oklahoma
73019, USA}
\date{\today}

\begin{abstract}
  The creation of artificial gauge fields in neutral ultracold atom
  systems has opened the possibility to study the effects
  of spin-orbit coupling terms in clean environments.
  This work considers the multi-channel scattering properties
  of two atoms confined by a wave guide in the presence of
  spin-orbit coupling terms within a K-matrix scattering framework.
  The tunability of resonances, induced by the interplay of the external
  wave guide geometry, the interactions, and the spin-orbit
  coupling terms, is demonstrated. 
Our results for the K-matrix elements as well as
partial and total reflection coefficients for
two identical
fermions interacting through a finite-range
interaction potential in the singlet channel only 
are compared with those obtained for a strictly 
one-dimensional effective low-energy Hamiltonian, which uses the
effective coupling constant derived in Zhang {\em{et al.}} 
[Scientific Reports {\bf{4}}, 1 (2014)] 
and 
Zhang {\em{et al.}} [Phys. Rev. A {\bf{88}}, 053605 (2013)]
as input.
In the regime where the effective Hamiltonian is applicable,
good agreement is obtained, provided the energy-dependence of the 
coupling constant is accounted for.
Our approach naturally
describes 
the energy regime in which the bands associated with excited transverse modes lie below
a subset of the bands associated with the lowest transverse modes.
  The threshold behavior is discussed and scattering observables are
  linked to bound state properties.
\end{abstract}


\maketitle

%%%%%%%%%%%%%%%%%%%%%%%%%%%%%%%%%%%%%%%%%%%%%%%%%%%%%%%%%%%%%%%%%%%%%%%%
%Introduction
%%%%%%%%%%%%%%%%%%%%%%%%%%%%%%%%%%%%%%%%%%%%%%%%%%%%%%%%%%%%%%%%%%%%%%%%
\section{Introduction}

Confinement-induced two-atom resonances occur when the length scale that
characterizes the outcome of the
low energy collision between two atoms
in free space is comparable to the size of the tight confinement 
length~\cite{Olshanii1998,PSW1D,school1}.
For a wave guide geometry with harmonic confinement in the 
$x$- and $y$-directions, the asymptotic even- or odd-$z$ solutions
along the wave guide direction ($z$-direction)
are the result of a multi-channel scattering calculation.
Since the energetically closed channels
are accessible during the collision process, the effective one-dimensional
even- and odd-$z$ coupling constants can be understood
as being renormalized by the energetically closed channels~\cite{Olshanii1998,Olshanii2,GrangerBlume}.
The K-matrix formalism ($\underline{K}$ is the reaction matrix) has been
shown to provide a transparent description of such
multi-channel problems~\cite{GrangerBlume,Panos_CE,Panos_multichannel}.

The present paper addresses what happens when the colliding atoms are additionally
feeling one-dimensional
spin-orbit coupling terms (equal mixture of Rashba
and Dresselhaus spin-orbit coupling)~\cite{Rashba1984,Dresselhaus1955,socreview1,socreview2}.
Among the various spin-orbit coupling schemes that have been
realized experimentally by now~\cite{Dalibardrev,YJLinSOC,socshaken,2dsoc1,2dsoc2,socOCT1,socOCT2}, 
the one-dimensional
spin-orbit coupling considered in this work is the most common.
Our work revisits the case where the spin-orbit coupling direction is
oriented along the wave guide axis~\cite{scientific_report,PRA_Zhang}.
A multi-channel K-matrix scattering theory that accounts for the
modification of the asymptotic solution due to the spin-orbit coupling
is developed.
Our theoretical framework is applied to two identical fermions 
with finite-range 
interaction in the singlet channel.
The theory is also applicable  to 
two identical bosons, and to distinguishable particles with spin-dependent
interactions.
It is found that even a relatively weak spin-orbit coupling strength
can lead to significant modifications of the resonance structure
that one would obtain in the absence of spin-orbit coupling,
thus providing an alternative route for controlling two-body resonances 
in a wave guide geometry.

The interplay between the external confinement and
the spin-orbit coupling terms has already been explored in two previous
publications~\cite{scientific_report,PRA_Zhang} for two identical fermions
interacting via zero-range interactions in the
singlet channel and vanishing interactions in the
triplet channels. 
Where comparisons can be made,
our results are in agreement with these earlier results.
The framework developed here is, however, more general
in that it is applicable to any type of interaction and any number of open channels.
The accomplishments of our work are:
\begin{itemize}
\item A general scattering framework applicable to two-particle
scattering in the presence of an external two-dimensional
harmonic trap and 
one-dimensional spin-orbit
coupling terms is developed.
\item The ``rotation approach'', 
introduced in Refs.~\cite{GBrotation,PengZhang}, 
is generalized to the wave guide problem and
used to interpret a subset of the results.
\item The effective one-dimensional coupling constant, derived
  in Ref.~\cite{scientific_report} in terms of a two-dimensional integral
  (see also Ref.~\cite{PRA_Zhang}),
  is found to be well approximated by the Hurwitz-Zeta 
function in certain parameter regimes. 
A physical picture of the energy-dependence of the 
Hurwitz-Zeta function
is provided.
\item The effective low-energy Hamiltonian is validated and K-matrix results
are also presented
in the energy regime, in which the effective 
low-energy Hamiltonian from the literature~\cite{scientific_report,PRA_Zhang}
is invalid.
\item The threshold laws in the vicinity of
various scattering thresholds are derived and interpreted.
\item The tunability of the scattering resonances is 
demonstrated and interpreted for two identical fermions.
\end{itemize}

The remainder of this paper is structured as follows.
Section~\ref{sec_setup} introduces the system Hamiltonian
and recasts, taking advantage of the symmetries of the system,
the associated Schr\"odinger equation in matrix form.
The 
scattering solutions of the matrix equation in the inner
region are obtained using the
generalized log-derivative
algorithm~\cite{GLD,SymmGLD}, which works when the usual second derivative operators
are complemented by first derivative
operators; in our case,
these arise from the spin-orbit coupling terms.
A discussion of the generalized log-derivative algorithm is relegated
to Appendix~\ref{appendix_algorithm}.
Section~\ref{sec_scatteringframework} discusses the asymptotic
solution that the inner solution is being matched to as well as 
the extraction of the physical K-matrix via channel elimination.
Taking a step back, Sec.~\ref{sec_rotation} introduces
an alternative approximate ``rotation approach'' that recasts
the coupled-channel problem in such a way that the first derivative operators 
are rotated away. This facilitates the use of standard algorithms such as the 
Johnson algorithm~\cite{Johnson}, 
thus vastly simplifying the numerics,
and provides a theoretical framework within which to interpret the
scattering results, at least in some parameter regimes.
The effective one-dimensional coupling 
constant~\cite{scientific_report,PRA_Zhang}, which enters
into the effective $4 \times 4$ low-energy Hamiltonian, is
introduced in Sec.~\ref{sec_1dcoupling} and the associated
threshold laws are analyzed.
Section~\ref{sec_results} applies the developed theory to
two identical fermions. 
Scattering quantities such as the partial
and total reflection coefficients are analyzed as a function of
the scattering energy.
To aid with the interpretation
of the scattering observables, we also calculate the 
corresponding two-fermion bound states.
Last, Sec.~\ref{sec_conclusion} provides a summary and an outlook.

%%%%%%%%%%%%%%%%%%%%%%%%%%%%%%%%%%%%%%%%%%%%%%%%%%%%%%%%%%%%%%%%%%%%%%%%
%scattering framework
%%%%%%%%%%%%%%%%%%%%%%%%%%%%%%%%%%%%%%%%%%%%%%%%%%%%%%%%%%%%%%%%%%%%%%%%
\section{Set-up of the problem}
\label{sec_setup}
We consider two identical point particles with mass $m$ 
that feel the single-particle Rashba-Dresselhaus spin-orbit
coupling $\hat{V}_{\text{so},j}$ ($j=1$ and 2)~\cite{socreview1,socreview2},
\begin{eqnarray}
  \label{eq_vso}
  \hat{V}_{\text{so},j}= 
\frac{\hbar k_{\text{so}} \hat{p}_{j,z}}{m} \hat{\sigma}_{j,z}
  +\frac{\hbar \Omega}{2} \hat{\sigma}_{j,x}
  + \frac{\hbar \delta }{2} \hat{\sigma}_{j,z},
\end{eqnarray}
as well as the single-particle harmonic potential 
$\hat{V}_{\text{trap},j}$ in the transverse
directions,
\begin{eqnarray}
  \label{eq_trap}
\hat{V}_{\text{trap},j}=\frac{1}{2} m \omega^2  \rho_j^2.
\end{eqnarray}
Here, $k_{\text{so}}$ is the strength of the spin-orbit coupling, $\Omega$ the Raman coupling
strength, $\delta$ the detuning, and $\omega$ the angular trapping frequency.
The position vectors of the particles are denoted by $\vec{r}_j$
(with components $x_j$, $y_j$, and $z_j$) and
the single-particle momentum operators by
$\hat{\vec{p}}_j$
(with components $\hat{p}_{j,x}$, $\hat{p}_{j,y}$, and $\hat{p}_{j,z}$).
The quantity $\rho_j$ is defined through $\rho_j^2 = x_j^2 + y_j^2$.
The spin-orbit coupling assumes that each atom can be considered as
containing two energy levels that form an effective spin-1/2 system,
described by the three Pauli matrices $\hat{\sigma}_{j,x}$,
$\hat{\sigma}_{j,y}$, and $\hat{\sigma}_{j,z}$.
This type of spin-orbit coupling is nowadays being realized
routinely in cold atom 
systems~\cite{YJLinSOC,MIT_fermionSOC,Zhang_fermionSOC,Peter_SOC,Yong_SOC,dipolarSOC,YtSOC}.
In addition to the single-particle potentials,
the particles feel a spin-dependent two-body interaction
potential $\hat{V}_{\text{int}}$,
\begin{eqnarray}
  \hat{V}_{\text{int}} =
  &&V_{S_0}(\vec{r})  |S_0\rangle \langle S_0|+
V_{T_{+1}}(\vec{r})|T_{+1}\rangle \langle T_{+1}| +
\nonumber \\
  &&
    V_{T_{-1}}(\vec{r}) |T_{-1}\rangle \langle T_{-1}|
+
  V_{T_0}(\vec{r})
   | T_0\rangle \langle T_0| .
\end{eqnarray}
For identical particles,
the interaction between the spin-up state of the first
atom and the spin-down
state of the second atom
is equal to
the interaction between the spin-down state of the first
atom and the spin-up
state of the second atom.
This implies: $V_{S_0}(\vec{r})=V_{T_0}(\vec{r})=V_{0}(\vec{r})$.
The interaction potential is written using the singlet-triplet
basis states
\begin{eqnarray}
\label{eq_spinstate1}
  |S_0 \rangle = \frac{1}{\sqrt{2}} 
\left( |\uparrow \downarrow \rangle - | \downarrow \uparrow \rangle \right),
\end{eqnarray}
\begin{eqnarray}
\label{eq_spinstate2}
  |T_{+1} \rangle =  |\uparrow \uparrow \rangle ,
\end{eqnarray}
\begin{eqnarray}
\label{eq_spinstate3}
  |T_{-1} \rangle =  |\downarrow \downarrow \rangle ,
\end{eqnarray}
and
\begin{eqnarray}
\label{eq_spinstate4}
  |T_0 \rangle = \frac{1}{\sqrt{2}} 
\left( |\uparrow \downarrow \rangle + | \downarrow \uparrow \rangle \right),
\end{eqnarray}
where ``$|\uparrow \rangle$'' and ``$| \downarrow \rangle$'' denote
the two internal states of the atoms.
Throughout this article, the potentials $V_{S_0}(\vec{r})$,
$V_{T_{+1}}(\vec{r})$,  $V_{T_{-1}}(\vec{r})$, and $V_{T_0}(\vec{r})$
($\vec{r}$ denotes the distance vector, $\vec{r}=\vec{r}_1-\vec{r}_2$)
are parametrized by spherically-symmetric
short-range potentials with range $r_0$.
The use of model interactions
like the Gaussian potential is justified if 
$r_0$ is notably smaller than the other length scales 
such as the transverse confinement length of the Hamiltonian.
Unless stated otherwise,
the ordering of the spin states given in
Eqs.~(\ref{eq_spinstate1})-(\ref{eq_spinstate4})
is used in the remainder of this article.

Denoting the kinetic energy $\hat{\vec{p}}_j^2/(2m)$
of the $j$-th particle by $\hat{T}_j$,
the system Hamiltonian $\hat{H}$ reads
\begin{eqnarray}
  \hat{H} = \left( \hat{T}_1  + \hat{T}_2 + \hat{V}_{\text{trap},1}   + 
\hat{V}_{\text{trap},2} \right) \hat{I}_1 \otimes \hat{I}_2 +
  \nonumber \\
  \hat{V}_{\text{so},1} \otimes \hat{I}_2 +   
\hat{I}_1 \otimes \hat{V}_{\text{so},2} 
+ 
\hat{V}_{\text{int}},
\end{eqnarray}
where $\hat{I}_j$ is the identity matrix of the spin Hilbert space of the $j$-th particle.
It can be readily checked that the $z$-component
$\hat{P}_z$ of the total momentum operator $\hat{\vec{P}}$,
\begin{eqnarray}
\hat{\vec{P}}=\hat{\vec{p}}_1 + \hat{\vec{p}}_2,
\end{eqnarray}
commutes with the total Hamiltonian $\hat{H}$~\cite{QZthesis}. 
This implies
that the expectation value $P_z$ of the operator
$\hat{P}_z$ is a good quantum number and that the
Schr\"odinger equation for the Hamiltonian $\hat{H}$ can be solved
separately for each $P_z$~\cite{scientific_report,PRA_Zhang}.
Using this, 
we separate $\hat{H}$ into the center-of-mass Hamiltonian $\hat{H}_{\text{cm}}$
and the relative Hamiltonian $\hat{H}_{\text{rel}}$~\cite{scientific_report,PRA_Zhang},
\begin{align}
\hat{H} = \hat{H}_{\text{cm}} + \hat{H}_{\text{rel}}.
\end{align}
We have
\begin{align}
  \hat{H}_{\text{cm}} = 
&\frac{{P}_z^2}{2M} \hat{I}_1 \otimes \hat{I}_2 + \nonumber \\
&\left( \frac{\hat{P}_x^2+\hat{P}_y^2}{2M} + \frac{1}{2} M \omega^2 (X^2 + Y^2) \right)
  \hat{I}_1 \otimes \hat{I}_2
  \end{align}
and
\begin{eqnarray}
\label{eq_hrel}
\hat{H}_\text{rel}=\hat{H}_{\text{ho}}+\hat{T}_{\text{rel},z}+
  \hat{V}_{\text{int}}+\hat{{V}}_{\text{so}} ,
\end{eqnarray}
where
\begin{eqnarray}
\hat{H}_{\text{ho}}=\left( \frac{\hat{p}_x^2 + \hat{p}_y^2}{2 \mu} +
  \frac{1}{2}\mu \omega_\bot^2 \rho^2 \right)
  \hat{I}_1\otimes\hat{I}_2 ,
  \end{eqnarray}
\begin{eqnarray}
  \hat{T}_{\text{rel},z}= 
\frac{\hat{p}_z^2}{2 \mu} \hat{I}_1 \otimes \hat{I}_2,
\end{eqnarray}
and
\begin{eqnarray}
  \label{eq_vso}
\hat{{V}}_{\text{so}} = 
&& 
\frac{\hbar k_{\text{so}}}{\mu}\hat{p}_z \hat{\Sigma}_z +
\frac{\hbar \Omega}{2}(\hat{\sigma}_{1,x}\otimes\hat{I}_2+\hat{I}_1\otimes \hat{\sigma}_{2,x}) + \nonumber \\
        && \frac{\hbar \tilde{\delta}}{2} 
        (\hat{\sigma}_{1,z}\otimes\hat{I}_2+\hat{I}_1\otimes \hat{\sigma}_{2,z})
\end{eqnarray}
with
\begin{eqnarray}
  \hat{\Sigma}_z = \frac{1}{2} \left(
  \hat{\sigma}_{1,z} \otimes \hat{I}_2- \hat{I}_1\otimes \hat{\sigma}_{2, z}
  \right)
\end{eqnarray}
and
\begin{eqnarray}
\label{eq_effectivedetuning}
    \frac{\hbar \tilde{\delta}}{2}=
    \frac{\hbar \delta}{2} + \frac{\hbar k_{\text{so}} P_z}{M}.
    \end{eqnarray}
Here, $M$ and $\mu$ denote the total mass and reduced mass,
respectively, and the center-of-mass vector $\vec{R}$ has
the components $X$, $Y$, and $Z$.
The relative momentum operator is denoted by $\hat{\vec{p}}$
and $\rho^2$ is defined through $x^2 + y^2$.
Note that $\hat{V}_{\text{so}}$ depends on the operator 
$\hat{p}_z$ but not on the operator $\hat{P}_z$
($\hat{P}_z$ is replaced by $P_z$).
Since the center-of-mass momentum $P_z$
can, according to Eq.~(\ref{eq_effectivedetuning}),
be interpreted as ``changing'' the physical detuning $\delta$,
we refer to $\tilde{\delta}$ as 
an effective or generalized detuning~\cite{GB3b}.
The center-of-mass Hamiltonian $\hat{H}_{\text{cm}}$ is identical to the
two-dimensional harmonic oscillator Hamiltonian for a particle
of mass $M$, with the
motion in the third dimension being governed by the free-particle Hamiltonian. 
The eigen states and eigen energies
of the corresponding Schr\"odinger equation can be written down readily.
In what follows, we focus  on the solutions to the
Schr\"odinger equation for the relative Hamiltonian
$\hat{H}_{\text{rel}}$.

To solve the relative Schr\"odinger equation, we
use that the $z$-component $\hat{l}_z$ of the orbital angular momentum operator $\vec{l}$
associated with the relative distance vector $\vec{r}$
commutes with $\hat{H}_{\text{rel}}$, i.e., $[\hat{l}_z,\hat{H}_{\text{rel}}]=0$.
This implies that we can determine the eigen states of $\hat{H}_{\text{rel}}$
separately for each $m_l$, where $m_l$ is the quantum number associated
with $\hat{l}_z$; $m_l$ takes the values $0, \pm 1,\cdots$.
We expand the eigen states $\Psi^{(m_l)}$ of $\hat{H}_{\text{rel}}$
for fixed $m_l$
as follows:
\begin{eqnarray}
  \label{eq_expand}
  \Psi^{(m_l)} = \sum_{n_{\rho},\chi}
  \phi_{n_{\rho},\chi}^{(m_l)}(z) \Phi_{n_{\rho}}^{(m_l)}(\rho) | \chi \rangle,
\end{eqnarray}
where the channel functions $ \Phi_{n_{\rho}}^{(m_l)}(\rho) | \chi \rangle$
are eigen states of $\hat{H}_{\text{ho}}$ with eigen energies
$\epsilon_{n_{\rho},m_l}$,
\begin{align}
\epsilon_{n_{\rho},m_l}=(2 n_{\rho} + |m_l|+1) \hbar \omega.
\end{align}
The radial quantum number 
$n_{\rho}$
takes the values $n_{\rho}=0,1,\cdots,n_{\text{max}}-1$,
where $n_{\text{max}}$ is the number of $\Phi_{n_{\rho}}^{(m_l)}$
included in the expansion.
The $\Phi_{n_{\rho}}^{(m_l)}(\rho)$ are, of course, just the
two-dimensional harmonic oscillator functions.
The index $\chi$ in Eq.~(\ref{eq_expand}) runs over the
spin functions: 
$|\chi \rangle=|S_0 \rangle $, $| T_{\pm 1} \rangle $, and $| T_0 \rangle$.
The ``weight functions''   $\phi_{n_{\rho},\chi}^{(m_l)}(z)$
are determined by plugging Eq.~(\ref{eq_expand})
into the relative Schr\"odinger equation $\hat{H}_{\text{rel}} \Psi^{(m_l)} = E \Psi^{(m_l)}$,
where $E$ denotes the relative scattering energy,
and solving the resulting set of coupled differential equations,
\begin{eqnarray}
  \label{eq_coupled_equations}
  \left( \epsilon_{n_{\rho},m_l}   +
  \hat{T}_{\text{rel},z} - E \right) 
\phi_{n_{\rho},\chi}^{(m_l)}(z) = \nonumber \\
  -\sum_{n_{\rho}'} {\cal{V}}_{\text{int}}^{n_{\rho},n_{\rho}',\chi}(z) \phi_{n_{\rho}',\chi}^{(m_l)}(z) -
  \sum_{\chi'} {{\cal{V}}}_{\text{so}} ^{\chi,\chi'}(\hat{p}_z) \phi_{n_{\rho},\chi'}^{(m_l)}(z).
  \end{eqnarray}
The matrix elements ${\cal{V}}_{\text{int}}^{n_{\rho}',n_{\rho},\chi}(z)$
are given by
\begin{eqnarray}
  {\cal{V}}_{\text{int}}^{n_{\rho}',n_{\rho},\chi}(z) =
  \langle \Phi_{n_{\rho}',\chi}^{(m_l)} | V_{\chi} | \Phi_{n_{\rho},\chi}^{(m_l)} \rangle,
\end{eqnarray}
where $V_{\chi}$ is equal to $V_{S_0}$, $V_{T_{\pm 1}}$, and 
$V_{T_0}$  for $|\chi\rangle = |S_0 \rangle$
$|T_{\pm 1} \rangle$, and $|T_0 \rangle$,
respectively.
The matrix elements 
${\cal{V}}_{\text{so}} ^{\chi',\chi} (\hat{p}_z)$ are given by
\begin{eqnarray}
  {{\cal{V}}}_{\text{so}}^{\chi',\chi}(\hat{p}_z) = 
\langle \chi' | \hat{{V}}_{\text{so}} | \chi \rangle .
  \end{eqnarray}
In deriving the coupled equations given in Eq.~(\ref{eq_coupled_equations}),
we used that the interaction potential 
$\hat{V}_{\text{int}}$ is diagonal in the
singlet-triplet basis (it couples different $n_{\rho}$) and that
the spin-orbit coupling term $\hat{{V}}_{\text{so}}$ is diagonal in the
harmonic oscillator basis states (it couples different $\chi$).

In practice, we solve the coupled equations
by rewriting Eq.~(\ref{eq_coupled_equations}) 
as a matrix of dimension $(4 n_{\text{max}}) \times (4 n_{\text{max}})$
(we denote the matrix by $\underline{H}_{\text{rel}}$)
acting on  the vector
\begin{eqnarray}
  \vec{\phi}^{(m_l)} =(\phi_{0,S_0}^{(m_l)}(z),\phi_{0,T_{+1}}^{(m_l)}(z),\cdots,\phi_{n_{\text{max}}-1,T_0}^{(m_l)}(z))^T.
\end{eqnarray}
Since the resulting equation $\underline{H}_{\text{rel}} \vec{\phi}^{(m_l)} = E \vec{\phi}^{(m_l)} $
has $4 n_{\text{max}}$ linearly independent solutions, we construct
the matrix $\underline{\phi}^{(m_l)}$, which contains the $j$-th eigen vector $\vec{\phi}^{(m_l)}$
in the $j$-th column,
and solve the resulting matrix equation
\begin{eqnarray}
\label{eq_SEmatrix}
  \underline{H}_{\text{rel}} \underline{\phi}^{(m_l)} = E  \underline{\phi}^{(m_l)}
\end{eqnarray}
numerically for relative energies $E$ equal to or greater than the
energy $E_{\text{th}}$ of the 
scattering threshold 
[see Eq.~(\ref{eq_minthreshold}) for an actual expression 
for $E_{\text{th}}$].
The details for enforcing the small-$|z|$
boundary condition and the algorithm for determining
the logarithmic derivative matrix $\underline{\cal{Y}}(z)$,
\begin{eqnarray}
\label{eq_logderivative}
\underline{\cal{Y}}(z) =
\frac{ d \underline{\phi}^{(m_l)}}{dz}
\left( \underline{\phi}^{(m_l)} \right) ^{-1},
\end{eqnarray}
are discussed in Appendix~\ref{appendix_algorithm}.

Our goal in this work is to determine the scattering solutions of the
relative Hamiltonian $\hat{H}_{\text{rel}}$ 
and to extract scattering observables from it.
To this end, the numerically obtained logarithmic
derivative
matrix $\underline{\cal{Y}}(z)$ needs to be matched to the
corresponding asymptotic large-$|z|$ solution, i.e., to the
solution obtained for $\hat{V}_{\text{int}}=0$.
Once the matching is done, the energetically closed
channels need to be eliminated.
The next section details these steps.

\section{Scattering framework}
\label{sec_scatteringframework}

\subsection{Asymptotic solution}
\label{sec_asymptotic}
The goal of this subsection is to determine the asymptotic
large-$|z|$
solutions, which are obtained by setting $\hat{V}_{\text{int}}$ to zero.
It follows from the discussion in Sec.~\ref{sec_setup}
that different $n_{\rho}$ channels are decoupled in the absence of
interactions.
This implies that the 
Hamiltonian matrix $\underline{H}_{\text{rel,ni}}$,
which is identical to $\underline{H}_{\text{rel}}$ except that
the matrix elements ${\cal{V}}_{\text{int}}^{n_{\rho}',n_{\rho},\chi}(z)$ are zero,
is block diagonal, with each fixed $n_{\rho}$-block having
dimension $4 \times 4$.
To write down the full solution, we solve one of the $4 \times 4$
blocks for fixed scattering energy $E$ and arbitrary
$n_{\rho}$.
In writing down the asymptotic
solution, we assume that $m_l$ is even and drop the
$m_l$ superscript for notational simplicity.
The changes required for odd $m_l$ are indicated in the text.

We denote the regular and irregular solutions
of the
$4 \times 4$ block by $\underline{f}_{n_{\rho}}$ 
and $\underline{g}_{n_{\rho}}$, respectively.
The 
$4 \times 4$ matrix $\underline{f}_{n_{\rho}}$
contains the eigen vector $\vec{f}_{n_{\rho}}^{(j)}$ in the $j$-th column
(and similarly for the irregular solution).
To obtain $\vec{f}_{n_{\rho}}^{(j)}$ and $\vec{g}_{n_{\rho}}^{(j)}$,
we make the ansatz
\begin{eqnarray}
\label{eq_fnrhoj}
  \vec{f}_{n_{\rho}}^{(j)} = \left(
  \begin{array}{c}
    a^{(j)}_1(k_{n_{\rho}}^{(j)}) {\cal{A}} (k_{n_{\rho}}^{(j)}|z| ) \\
    \imath a^{(j)}_2(k_{n_{\rho}}^{(j)}) {\cal{B}} (k_{n_{\rho}}^{(j)}|z| ) \\
    \imath a^{(j)}_3(k_{n_{\rho}}^{(j)}) {\cal{B}} (k_{n_{\rho}}^{(j)}|z| ) \\
    \imath a^{(j)}_4(k_{n_{\rho}}^{(j)}) {\cal{B}} (k_{n_{\rho}}^{(j)}|z| ) \\
  \end{array}
  \right)
  \Phi_{n_{\rho}}(\rho)
  \end{eqnarray}
and 
\begin{align}
\label{eq_gnrhoj}
  \vec{g}_{n_{\rho}}^{(j)} = \left(
  \begin{array}{c}
    a^{(j)}_1(k_{n_{\rho}}^{(j)}) {\cal{C}} (k_{n_{\rho}}^{(j)}|z| ) \\
    \imath a^{(j)}_2(k_{n_{\rho}}^{(j)}) {\cal{D}} (k_{n_{\rho}}^{(j)}|z| ) \\
    \imath a^{(j)}_3(k_{n_{\rho}}^{(j)}) {\cal{D}} (k_{n_{\rho}}^{(j)}|z| ) \\
    \imath a^{(j)}_4(k_{n_{\rho}}^{(j)}) {\cal{D}} (k_{n_{\rho}}^{(j)}|z| ) \\
  \end{array}
  \right)
  \Phi_{n_{\rho}}(\rho),
  \end{align}
where
\begin{align}
  {\cal{A}} (k_{n_{\rho}}^{(j)}|z| ) =
  \Bigg\{ \begin{array}{c}
    \cos (k_{n_{\rho}}^{(j)}|z| ) \mbox{ for FF } (m_l \mbox{ even}) \\
    \mbox{sign}(z)  \sin (k_{n_{\rho}}^{(j)}|z| ) \mbox{ for BB } (m_l \mbox{ even}) 
  \end{array},
  \end{align}
\begin{align}
  {\cal{B}} (k_{n_{\rho}}^{(j)}|z| ) =
  \Bigg\{ \begin{array}{c}
    \mbox{sign}(z)\sin (k_{n_{\rho}}^{(j)}|z| ) \mbox{ for FF } (m_l \mbox{ even})\\
      -\cos (k_{n_{\rho}}^{(j)}|z| ) \mbox{ for BB } (m_l \mbox{ even}) 
  \end{array},
\end{align}
\begin{align}
  {\cal{C}} (k_{n_{\rho}}^{(j)}|z| ) =
  \Bigg\{ \begin{array}{c}
    \sin (k_{n_{\rho}}^{(j)}|z| ) \mbox{ for FF } (m_l \mbox{ even})\\
    -\mbox{sign}(z)  \cos (k_{n_{\rho}}^{(j)}|z| ) \mbox{ for BB } (m_l \mbox{ even}) 
  \end{array},
\end{align}
and
\begin{align}
\label{eq_cald}
  {\cal{D}} (k_{n_{\rho}}^{(j)}|z| ) =
  \Bigg\{ \begin{array}{c}
    -\mbox{sign}(z) \cos (k_{n_{\rho}}^{(j)}|z| ) \mbox{ for FF } (m_l \mbox{ even})\\
      -\sin (k_{n_{\rho}}^{(j)}|z| ) \mbox{ for BB } (m_l \mbox{ even}) 
  \end{array}.
  \end{align}
In Eqs.~(\ref{eq_fnrhoj})-(\ref{eq_cald}), 
the spatial parts are chosen such that the components of
$\vec{f}_{n_{\rho}}^{(j)}$ and $\vec{g}_{n_{\rho}}^{(j)}$ have the correct symmetry
for two identical fermions (FF) and two identical bosons (BB). Specifically,
the spin singlet state is anti-symmetric under the exchange of
the spins of the first and second particles. Thus,
the corresponding spatial part for two identical fermions
has to be symmetric under the exchange
of the spatial degrees of freedom
while that for two identical bosons has to be anti-symmetric.
Since $m_l$ is assumed to be even,
the functions $\Phi_{n_{\rho}}$ are unchanged when exchanging the spatial coordinates
of the particles. This implies that the functions ${\cal{A}}$ and ${\cal{C}}$
have to be even for two identical fermions and odd for two identical bosons.
The argument for the spin triplet components follows the same logic.

The $k_{n_{\rho}}^{(j)}$ are defined in terms of the scattering energy $E$,
\begin{eqnarray}
\label{eq_scattering_escatt}
  E = \epsilon_{n_{\rho},m_l}+ E_z^{(j)}(\hbar k_{n_{\rho}}^{(j)}),
\end{eqnarray}
where the $E_z^{(j)}$ are
the four relative free-particle dispersion curves.
The dispersion curves are obtained by solving a
quartic equation in $E_z^{(j)}$.
One finds~\cite{soc1dpaper}
\begin{eqnarray}
\label{eq_E1}
E^{(1/2)}_z(p_z)=\frac{p_z^2}{m} \pm \sqrt{ 2 a - 2 \sqrt{a^2 -\frac{p_z^2}{m} E_{\text{so} }
E_{\tilde{\delta}}^2}}
\end{eqnarray}
and
\begin{eqnarray}
\label{eq_E3}
E^{(3/4)}_z(p_z)=\frac{p_z^2}{m} \pm \sqrt{ 2 a + 2 \sqrt{a^2 -\frac{p_z^2}{m} E_{\text{so} }
E_{\tilde{\delta}}^2}},
\end{eqnarray}
where
\begin{align}
  \label{eq_aux_a}
a = \frac{p_z^2}{m} E_{\text{so}} + \frac{1}{4} \left(
E_{\Omega}^2 + E_{\tilde{\delta}}^2 \right).
\end{align}
Here, we defined
\begin{align}
  E_\Omega=\hbar\Omega,
\end{align}
\begin{align}
  E_\text{so}=\frac{\hbar^2 k_{\text{so}}^2}{m},
  \end{align}
and 
\begin{align}
  E_{\tilde{\delta}}= \hbar \tilde{\delta}.
  \end{align}
The plus and minus signs in Eq.~(\ref{eq_E1}) are for 
$E^{(1)}_z(p_z)$ and $E^{(2)}_z(p_z)$, respectively.
The plus and minus signs in Eq.~(\ref{eq_E3}) are for 
$E^{(3)}_z(p_z)$ and $E^{(4)}_z(p_z)$, respectively.
Explicit expressions for the vectors $\vec{a}^{(j)}$ 
for $\tilde{\delta}=0$
are
reported in Appendix~\ref{appendix_asymptotics}.

For odd $m_l$, the asymptotic solutions for 
two identical fermions given above become the
solutions for two identical bosons, and vice versa.
For two distinguishable particles, no symmetry constraints exist,
implying that the ``bosonic'' and ``fermionic'' solutions need 
to be combined.

Having the regular $4 \times 4$
matrix solutions $\underline{f}_{n_{\rho}}$,
we define the matrix $\underline{f}$ through
\begin{eqnarray}
  \underline{f}=\left(
  \begin{array}{cccc}
    \underline{f}_0 & 0 & \cdots & 0 \\
    0 & \underline{f}_1 &  &  \\
    \vdots &  & \ddots &  \\
    0 &  &  & \underline{f}_{n_{\text{max}}-1}
    \end{array}
  \right) .
\end{eqnarray}
The matrix $\underline{g}$ is defined analogously.
With these definitions, the asymptotic large-$|z|$ solution
$\underline{\Psi}_{\text{out}}$ reads
\begin{eqnarray}
  \underline{\Psi}_{\text{out}}
=
  \underline{f} - \underline{g} \underline{K} . 
  \end{eqnarray}
The K-matrix is obtained by equating the inner solution $\underline{\Psi}$
and the outer solution $\underline{\Psi}_{\text{out}}$ as well as their derivatives
with respect to $z$ at $z=z_{\text{max}}$, where $z_{\text{max}}$ is chosen such that the
inner solution $\underline{\Psi}$ has reached its asymptotic behavior, i.e.,
such that the phase accumulation due to the interaction potential has
reached a converged value.
In terms of the logarithmic derivative matrix
$\underline{\cal{Y}}(z)$ at $z_{\text{max}}$ 
[see Eq.~(\ref{eq_logderivative})],
the K-matrix can be written as~\cite{GBrotation}
\begin{align}
\label{eq_kfromy}
\underline{K}=\bigg[ \underline{\cal{Y}}(z) \underline{g}(z)-
\frac{d \underline{g}(z)}{dz} \bigg]^{-1}
 \bigg[  \underline{\cal{Y}}(z) \underline{f}(z)-
\frac{d \underline{f}(z)}{dz} \bigg]_{z=z_\text{max}}.
\end{align}

\begin{figure}[h]
\begin{center}
\includegraphics[width=3.2in]{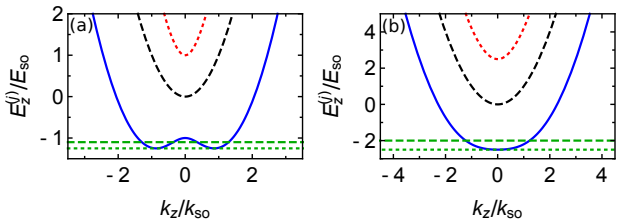}
  \caption{(color online)
    Non-interacting relative dispersion curves $E_z^{(j)}$ as
    a function of $k_z/k_{\text{so}}$ for $\tilde{\delta}=0$
    and (a) $E_{\Omega} =  E_{\text{so}}$ (double-minimum regime) and
    (b) $E_{\Omega} = 5 E_{\text{so}}/2$ (single-minimum regime).
    The green
dotted horizontal lines depict $E_{\text{th}}$ [Eq.~(\ref{eq_thresholdenergy})].
    The green
dashed horizontal lines depict a scattering energy $E$ for which
    the number of energetically open channels,
    assuming $2 \hbar \omega$ is larger than
    the difference between $E$ and $E_{\text{th}}$,
    is (a) two and (b) one.
The discussion of the threshold laws in Sec.~\ref{sec_1dcoupling_soc}
uses the ``Roman labeled'' energies $E_{I}$, $E_{II}$,
$E_{II'}$, $E_{III}$, and $E_{IV}$.
The energy $E_{I}$ is equal to $E_{\text{th}}$ for the double-minimum case.
 The energy $E_{II}$ is equal to $E_{\text{th}}$ for the single-minimum case.
The energy $E_{II'}$ is given by the local maximum of the
blue 
solid line in the double-minimum regime, $E_{II'}=\hbar \omega - E_{\Omega}$.
The minima of the black dashed 
and red dotted curves have the energies $E_{III}=\hbar \omega$
and 
$E_{IV}=\hbar \omega + E_{\Omega}$, respectively.
    Assuming $2 \hbar \omega$ is larger than
    $E_{\text{so}}+E_{\Omega}+E_\Omega^2/(4E_{\text{so}})$, 
the number of energetically open channels changes from
two to one at $E_{II'}$,
from one to three at $E_{III}$, and 
from three to four at $E_{IV}$.
    }
\label{fig1}
\end{center}
\end{figure}

Figures~\ref{fig1}(a) and \ref{fig1}(b) show
the relative dispersion curves $E_z^{(j)}(p_z)$
as a function of the relative wave vector
$k_z$ in the $z$-direction, which is defined as $k_z = p_z/ \hbar$,
for $\tilde{\delta}=0$
in the double-minimum regime ($E_{\Omega} = E_{\text{so}}$)
and in the single-minimum regime  ($E_{\Omega} = 5 E_{\text{so}}/2$),
respectively.
For $\tilde{\delta}=0$, 
the transition from the double-minimum to the single-minimum
regime occurs at $\hbar \Omega_*=2 E_{\text{so}}$.
The minimum of the lowest dispersion curve defines the
scattering threshold $E_{\text{th}}$,
\begin{align}
\label{eq_minthreshold}
E_{\text{th}} = \min_{p_z,j} \left( E_z^{(j)}(p_z) + \hbar \omega \right).
\end{align}
For $\tilde{\delta}=0$, one finds
\begin{equation}
\label{eq_thresholdenergy}
E_{\text{th}}=
  \begin{cases}
    \hbar\omega-E_{\text{so}} -
    \frac{(E_{\Omega})^2}{4 E_{\text{so}}} & \mbox{ for }  \Omega<\Omega_*\\
    \hbar\omega-E_{\Omega} & \mbox{ for }  \Omega>\Omega_*
  \end{cases}.
\end{equation}
Scattering solutions are obtained for energies $E$
equal to or greater than $E_{\text{th}}$.
Inspection of Eqs.~(\ref{eq_scattering_escatt})-(\ref{eq_aux_a})
shows that the $k_{n_{\rho}}^{(j)}$ can be imaginary
(whether or not they are imaginary depends on the values of $E$, $E_{\text{so}}$,
$E_{\Omega}$, $E_{\tilde{\delta}}$, and $\hbar \omega$).
If a $k_{n_{\rho}}^{(j)}$ is imaginary, the solution
blows up exponentially at large $|z|$ in the respective channel.
Physically, the channel is energetically closed and needs to be
eliminated.
The next subsection illustrates how the energetically
closed channels are eliminated to obtain the physical 
K-matrix $\underline{K}_{\text{phys}}$.

\begin{figure}[h]
\begin{center}
\includegraphics[width=3.2in]{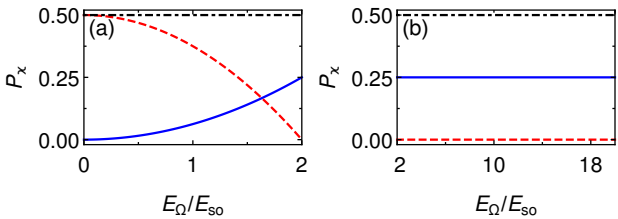}
  \caption{(color online)
Spin composition of the state corresponding to
the lowest
scattering threshold
for $\tilde{\delta}=0$ 
as a function of $E_{\Omega}/E_{\text{so}}$
for (a) $E_{\Omega} < 2 E_{\text{so}}$ (double-minimum regime)
and
(b) $E_{\Omega} > 2 E_{\text{so}}$ (single-minimum regime).
The probabilities $P_{\chi}$ to be in the $|S_0 \rangle$,
$|T_0 \rangle$, and $|T_{+1} \rangle$
channels are shown by red dashed, black dash-dotted, 
and blue solid lines,
respectively.
The probability to be in the $|T_{-1} \rangle$ channel
(not shown) is equal to the
probability to be in the $|T_{+1} \rangle$ channel.
    }
\label{fig_threshold}
\end{center}
\end{figure}

Figure~\ref{fig_threshold}
shows
the probability $P_{\chi}$
that the state corresponding to the lowest
scattering threshold
for $\tilde{\delta}=0$
is in the spin channel $|\chi \rangle$
as a function of $E_{\Omega}/E_{\text{so}}$.
As $\Omega$ increases from zero to $\Omega_*$ [Fig.~\ref{fig_threshold}(a)], 
the spin composition 
changes quite a bit. For $\Omega > \Omega_*$ 
[Fig.~\ref{fig_threshold}(b)], in contrast,
the spin composition of the scattering threshold is constant.
For infinitesimally small $\Omega$, the state corresponding to the
lowest scattering threshold
contains predominantly $|S_0 \rangle$ and $|T_0 \rangle$
admixtures.
The $|S_0 \rangle$ contribution decreases to zero
as $\Omega$ reaches $\Omega_*$ and remains zero for $\Omega > \Omega_*$. 
The spin-composition of the state at the lowest scattering
threshold 
is used in Sec.~\ref{sec_results}
to interpret the scattering observables.
The second lowest scattering threshold
is two-fold degenerate.
Importantly, one of the
associated states is a pure 
$|S_0 \rangle$ state for all $\Omega$
and the other is a superposition of the $|T_{+1} \rangle$
and $|T_{-1} \rangle$ channels.
The fact that one of the two second lowest threshold
states corresponds to the $|S_0 \rangle$ channel plays an important role
in understanding the resonance structure for $\Omega > \Omega_*$.

\subsection{Physical K-matrix}
\label{sec_physical}

To perform the channel 
elimination~\cite{Seatonrev,QDT1,QDT2,Chrisrev,GrangerBlume,Panos_multichannel,Panos_CE}, 
we partition
$\underline{\Psi}_{\text{out}}$, $\underline{f}$, $\underline{g}$, and 
$\underline{K}$ into
four blocks: the open-open (``oo''), open-closed
(``oc''), closed-open (``co''), and
closed-closed (``cc'') blocks.
To perform this partitioning, the
columns and rows of the matrices defined in the previous section
may need to be reordered.
If we consider a scattering energy $E$ that is just slightly
above the lowest scattering threshold
with energy $E_{\text{th}}$ and if we consider 
spin-orbit coupling parameters such that the single-minimum
regime is realized, then there is one open channel
and the open-open K-matrix corresponds to a $1 \times 1$ matrix.
If, on the other hand, we consider the double-minimum regime with $E$
just above $E_{\text{th}}$
and $\tilde{\delta}=0$, then there are two open channels and the open-open
K-matrix corresponds to a $2 \times 2$ matrix.
Formally, we write
\begin{align}
\label{eq_fmgK}
\begin{pmatrix}
 \underline{\Psi}^\text{oo}_{\text{out}} &  \underline{\Psi}^\text{oc}_{\text{out}}   \\
  \underline{\Psi}^\text{co}_{\text{out}}  &   \underline{\Psi}^\text{cc}_{\text{out}}  
\end{pmatrix}
=\begin{pmatrix}
 \underline{f}^\text{o} &  \underline{0}   \\
  \underline{0}  &   \underline{f}^\text{c}  
\end{pmatrix}-
\begin{pmatrix}
 \underline{g}^\text{o} &  \underline{0}  \\
 \underline{0} &   \underline{g}^\text{c}  
\end{pmatrix}
\begin{pmatrix}
 \underline{K}^{\text{oo}} &  \underline{K}^{\text{oc}}   \\
 \underline{K}^{\text{co}}  &  \underline{K}^{\text{cc}}    
\end{pmatrix},
\end{align} 
where the matrices with superscripts ``oo'', ``cc'', ``co'',
and ``oc'' have
dimensions $N^{\text{o}} \times N^{\text{o}}$,
$N^{\text{c}} \times N^{\text{c}}$,
$N^{\text{c}} \times N^{\text{o}}$,
and
$N^{\text{o}} \times N^{\text{c}}$
($N^{\text{o}}+N^{\text{c}}= 4 n_{\text{max}}$).

The idea is~\cite{Seatonrev,Chrisrev,QDT1,QDT2} 
to construct a linear combination of 
the asymptotic solution, Eq.~(\ref{eq_fmgK}), so 
that the divergent parts are removed. 
Following Refs.~\cite{Seatonrev,Chrisrev,QDT1,QDT2,Panos_CE}, 
we write the coefficient matrix that ``rearranges'' the
asymptotic solution as $(\underline{I},\underline{Y}^{\text{co}})^T$,
where $\underline{I}$ is the identity matrix of size $N^{\text{o}} \times N^{\text{o}}$.
Acting with both sides of Eq.~(\ref{eq_fmgK}) onto
$(\underline{I},\underline{Y}^{\text{co}})^T$,
we obtain
\begin{align}
\label{eq_lc}
\begin{pmatrix}
 \underline{\Psi}^\text{oo}_{\text{out}} +  \underline{\Psi}^\text{oc}_{\text{out}} \underline{Y}^\text{co}   \\
  \underline{\Psi}^\text{co}_{\text{out}}  +   \underline{\Psi}^\text{cc}_{\text{out}}  \underline{Y}^\text{co} 
\end{pmatrix} 
= \nonumber \\
\begin{pmatrix}
 \underline{f}^\text{o} -\underline{g}^\text{o} 
\underline{K}^\text{oo} & -\underline{g}^\text{o} \underline{K}^\text{oc}   \\
 -\underline{g}^\text{c} \underline{K}^\text{co} & 
\underline{f}^\text{c}-\underline{g}^\text{c} \underline{K}^\text{cc}  
\end{pmatrix}
 \begin{pmatrix}
\underline{I} \\
\underline{Y}^\text{co}
\end{pmatrix}.
\end{align} 
To remove the diverging part, $\underline{Y}^\text{co}$ needs to satisfy 
\begin{align}
\label{eq_findYco}
-\underline{g}^\text{c} \underline{K}^\text{co} + 
\left( \underline{f}^\text{c}-\underline{g}^\text{c} \underline{K}^\text{cc}
\right)  
\underline{Y}^\text{co}=0.
\end{align}
Solving for $\underline{Y}^{\text{co}}$ and inserting the result
into the first line of Eq.~(\ref{eq_lc}), we find
\begin{align}
 \label{eq_der1}
 &\underline{\Psi}^\text{oo}_{\text{out}} +  
\underline{\Psi}^\text{oc}_{\text{out}} \underline{Y}^\text{co} \nonumber \\
 &= 
\underline{f}^\text{o}- \underline{g}^\text{o} \underline{K}^\text{oo} -
\underline{g}^\text{o} \underline{K}^\text{oc} 
\left[({\underline{g}^\text{c}})^{-1} \underline{f}^\text{c}-\underline{K}^\text{cc}
\right]^{-1} \underline{K}^{\text{co}}.
\end{align}
Using that
$({\underline{g}^\text{c}})^{-1} \underline{f}^\text{c}=
- \imath \underline{I}$~\cite{Seatonrev,Chrisrev,QDT1,QDT2,Panos_CE},
we obtain the final expression,
\begin{align}
 \label{eq_der2}
 \underline{\Psi}^\text{oo}_{\text{out}} +  
\underline{\Psi}^\text{oc}_{\text{out}} \underline{Y}^\text{co} =
\underline{f}^{\text{o}} - \underline{g}^{\text{o}} \underline{K}_{\text{phys}},
\end{align}
where the
physical K-matrix $\underline{K}_{\text{phys}}$ is defined 
as~\cite{Chrisrev,GrangerBlume,Panos_CE}
\begin{align}
\label{eq_Kphys}
\underline{K}_{\text{phys}}=
\underline{K}^\text{oo}+ \imath \underline{K}^\text{oc}
(\underline{I}- \imath \underline{K}^\text{cc})^{-1} \underline{K}^\text{co}.
\end{align}
The first term on the right hand side
of Eq.~(\ref{eq_Kphys}) is the ``usual'' term, 
which describes particles entering and leaving in the open channel(s). 
The second term incorporates ``higher-order processes'', which describe 
particles entering in the open channel,
transitioning to intermediate closed channels, and leaving in the open channel.

In addition to the K-matrix, we consider the scattering amplitude matrix
$\underline{{\cal{F}}}$
(note that the scattering amplitude matrix is usually denoted by $\underline{f}$;
however, we use $\underline{{\cal{F}}}$ instead since the symbol $\underline{f}$
is used to denote the regular solution).
The scattering amplitude matrix $\underline{\cal{F}}$,
\begin{align}
\label{eq_scattering_amplitude}
\underline{\cal{F}} = \imath \underline{K}_{\text{phys}} 
\left(
\underline{I} - \imath
\underline{K}_{\text{phys}}
\right)^{-1},
\end{align}
at fixed scattering energy $E$
(the energy dependence is not indicated explicitly)
defines the matrix elements
${\cal{T}}_{st}$ and ${\cal{R}}_{st}$ of the transmission
coefficient  matrix $\underline{\cal{T}}$
and the
reflection coefficient
matrix $\underline{\cal{R}}$, respectively,
\begin{align}
\label{eq_calt}
{\cal{T}}_{st}=| \delta_{st} + {\cal{F}}_{st}|^2
\end{align}
and
\begin{align}
\label{eq_calr}
{\cal{R}}_{st}=|{\cal{F}}_{st}|^2.
\end{align}
In Eqs.~(\ref{eq_calt})-(\ref{eq_calr}),
the second subscript denotes the incoming channel and the first
subscript the outgoing channel.
This implies that the transmission coefficient
${\cal{T}}_t$, which quantifies the fraction of transmitted flux,
provided the incoming flux is located in
channel $t$, is given by
\begin{align}
\label{eq_calt_partial}
{\cal{T}}_t = \sum_{s \, \in \, \text{open}} {\cal{T}}_{st}.
\end{align}
Similarly,
the reflection coefficient
${\cal{R}}_t$, which quantifies the fraction of reflected flux,
provided the incoming flux is located in 
channel $t$, is given by
\begin{align}
\label{eq_calr_partial}
{\cal{R}}_t = \sum_{s \, \in \, \text{open}} {\cal{R}}_{st}.
\end{align}
The total transmission and reflection coefficients
$\cal{T}$ and $\cal{R}$ are defined by
\begin{align}
\label{eq_calt_total}
{\cal{T}}=\sum_{t \, \in \, \text{open}} {\cal{T}}_t
\end{align}
and
\begin{align}
\label{eq_calr_total}
{\cal{R}}=\sum_{t \, \in \, \text{open}} {\cal{R}}_t,
\end{align}
respectively.
In Eqs.~(\ref{eq_calt_partial})-(\ref{eq_calr_total})
the sum extends over the $N^{\text{o}}$ energetically open channels.
As required by flux conservation, the above definitions
are consistent with the identity
\begin{align}
{\cal{T}} + {\cal{R}}=N^{\text{o}}.
\end{align}
If $\underline{K}_{\text{phys}}$ has only one non-vanishing eigen value
(denoted by ${K}_{\text{phys}}^{(1)}$), then
${\cal{R}}$ is given by
\begin{align}
\label{eq_connect_k_and_r}
{\cal{R}}=\frac{(K_{\text{phys}}^{(1)})^2}{1+(K_{\text{phys}}^{(1)})^2}.
\end{align}

\section{Approximate ``rotation approach''}
\label{sec_rotation}
This section introduces an alternative but approximate scheme,
referred to as ``rotation approach''~\cite{GBrotation}, to
obtain the scattering observables numerically. Compared to the full
coupled-channel treatment, the
rotation approach is numerically more efficient
and does not require the implementation of a propagator customized
for the treatment of systems with spin-orbit coupling
(Hamiltonian that contains terms proportional to 
$\hat{p}_{z}^2$ and $\hat{p}_z$).
While it is not fully clear 
how to estimate
the accuracy of the rotation approach {\em{a priori}},
numerical tests show that it works quite accurately for a wide
range of spin-orbit coupling parameters.
In addition to simplifying the numerics, the rotation approach
can also be used to make back-of-the-envelope type estimates of the
expected resonance structure based on the knowledge of the system
without spin-orbit coupling, at least for a subset of the parameter 
space.

Just as the full
coupled-channel approach, the rotation approach divides the space along $z$
into an inner and an outer region. The full Hamiltonian in the inner region
is replaced
by a rotated Hamiltonian.
The rotation needs to be undone when writing out the matching condition.

The idea is to define the rotated Hamiltonian
$\hat{H}_{\text{rel}}^{R}$ in terms of a rotation 
operator $\hat{R}$~\cite{GBrotation,PengZhang},
\begin{align}
\hat{H}_{\text{rel}}^{{R}}=\hat{R}^{-1}\hat{H}_{\text{rel}} \hat{R},
\end{align}
such that $\hat{H}_{\text{rel}}^{R}$ contains a second derivative with 
respect to $z$ but not a first derivative with respect to $z$.
Choosing (inspired by Refs.~\cite{GBrotation,PengZhang}) 
\begin{equation}
\hat{R}=\exp(-\imath k_{\text{so}} \hat{\Sigma}_z z)
\end{equation}
and assuming $V_{S_0}(\vec{r})=V_{T_0}(\vec{r})=V_0(\vec{r})$,
we find
\begin{align}
\label{eq_hrot}
\hat{H}_{\text{rel}}^{{R}}=
\hat{H}_{\text{ho}} + \hat{T}_{\text{rel},z} + \hat{V}_{\text{int}}+
\hat{{V}}_{\text{so}}^{R},
\end{align}
where
\begin{align}
\label{eq_vso_rot}
\hat{{V}}_\text{so}^{R}=&
- E_\text{so} \hat{\Sigma}_z^2 \nonumber \\
                       &+\frac{\hbar \Omega}{2}\bigg[\hat{\sigma}_{1,x}\otimes\hat{I}_2 \cos(k_{\text{so}}z)-\hat{\sigma}_{1,y}\otimes\hat{I}_2 \sin(k_{\text{so}}z)\bigg] \nonumber \\
                       &+\frac{\hbar \Omega}{2}\bigg[\hat{I}_1\otimes \hat{\sigma}_{2,x}\cos(k_{\text{so}}z)-\hat{I}_1\otimes \hat{\sigma}_{2,y}\sin(k_{\text{so}}z)\bigg]                        \nonumber \\
                       &  +\frac{\hbar \tilde{\delta}}{2}(\hat{\sigma}_{1,z}\otimes\hat{I}_2+\hat{I}_1\otimes \hat{\sigma}_{2,z})
.
\end{align}
Equations~(\ref{eq_hrot}) 
and (\ref{eq_vso_rot}) show that the term 
linear in $\hat{p}_z$ is, indeed,
absent. The simplification of the derivative terms comes ``at a price'', however.
The rotation or gauge transformation
introduces a new term [first term on the right-hand-side of
Eq.~(\ref{eq_vso_rot})] as well as a spatially oscillating 
Raman coupling strength $\Omega$.
So far, no approximations have been made, i.e.,
the rotated Hamiltonian in Eq.~(\ref{eq_hrot}) is equivalent
to the original Hamiltonian $\hat{H}_{\text{rel}}$ 
given in Eq.~(\ref{eq_hrel}).

Assuming that
$k_{\text{so}}|z|$ is small and Taylor expanding the $\sin$ and $\cos$ terms to leading order,
we find the following small-$|z|$ expression for the spin-orbit
coupling term,
\begin{align}
\label{eq_vso_smallz_sr}
\hat{{V}}_{\text{so}}^{R,\text{sr}} = &
 -E_\text{so} \hat{\Sigma}_z^2 +
\frac{\hbar \Omega}{2}(\hat{\sigma}_{1,x}\otimes\hat{I}_2+\hat{I}_1\otimes \hat{\sigma}_{2,x}) + \nonumber \\
        & \frac{\hbar \tilde{\delta}}{2} 
        (\hat{\sigma}_{1,z}\otimes\hat{I}_2+\hat{I}_1\otimes \hat{\sigma}_{2,z}),
\end{align}
where the superscript ``sr'' indicates that this
expression is only valid for sufficiently small $k_{\text{so}} |z|$.
The next order correction is proportional to $\Omega k_{\text{so}} |z|$.
The premise is that the phase accumulation has reached its asymptotic value,
at least to a very good approximation, before the next-order terms in the
expansions of the oscillating $\Omega$ terms become important.
We denote the corresponding small-$|z|$ Hamiltonian by 
$\hat{H}_{\text{rel}}^{R,\text{sr}}$.
Since
$\hat{H}_{\text{rel}}^{R,\text{sr}}$
does not contain any first derivative terms with respect to $z$,
the corresponding Schr\"odinger equation can be
propagated using any ``standard''  propagator.
Denoting the resulting logarithmic derivative matrix by
$\underline{\cal{Y}}^{R,\text{sr}}(z)$, the
approximate
logarithmic derivative matrix
in the singlet-triplet basis,
obtained within the rotation approach, 
reads
(see 
Ref.~\cite{GBrotation} for an analogous derivation)
\begin{align}
\label{eq_ysoc1}
\underline{\cal{Y}}^{\text{sr}}(z) =&
\left(\frac{d \underline{{R}}(z)}{dz} \right) \,
\left(\underline{{R}}(z)\right)^{-1}+ \nonumber \\
&\underline{{R}}(z) \,  \underline{\cal{Y}}^{R,\text{sr}}(z) \,
\left(\underline{{R}}(z) 
\right)^{-1},
\end{align}
where $\underline{{R}}(z)$ is the matrix representation
of $\hat{R}$ in the singlet-triplet basis
(see Appendix~\ref{appendix_rotation}).

Expressed in the spin basis $\{ |R_j \rangle \}$
(see Appendix~\ref{appendix_rotation}),
$\hat{{V}}_{\text{so}}^{R,\text{sr}}$ is not diagonal.
In the special case that the interactions in the singlet and
triplet channels are all equal
[$V_{0}(\vec{r})=V_{T_{+1}}(\vec{r})=V_{T_{-1}}(\vec{r})$],
$\hat{{V}}_{\text{so}}^{R,\text{sr}}$ can be diagonalized
by applying another
transformation.
Defining
\begin{align}
\hat{H}_{\text{rel}}^{UR,\text{sr}}= 
U^{\dagger} \hat{H}_{\text{rel}}^{R,\text{sr}} U, 
\end{align}
the resulting small-$|z|$ Hamiltonian $\hat{H}_{\text{rel}}^{UR,\text{sr}}$
reads
\begin{eqnarray}
\label{eq_hrel_ur}
\hat{H}_{\text{rel}}^{UR,\text{sr}}=
\hat{H}_{\text{ho}} + \hat{T}_{\text{rel},z} + \hat{V}_{\text{int}} + 
\sum_{j=1}^4 \epsilon_j |D_j \rangle \langle D_j |,
\end{eqnarray}
where the energy shifts $\epsilon_j$ are
determined by the solutions to the equation
\begin{align}
  \epsilon_j^4 + 2 E_{\text{so}} \epsilon_j^3
  + (E_{\text{so}}^2 - E_{\Omega}^2 - E_{\tilde{\delta}}^2 ) \epsilon_j^2
  - \nonumber \\
  (E_{\Omega}^2 + 2 E_{\tilde{\delta}}^2) E_{\text{so}} \epsilon_j
  - E_{\text{so}}^2 E_{\tilde{\delta}}^2 =0.
  \end{align}
For $\tilde{\delta}=0$, one finds
\begin{align}
\label{eq_epsilon1}
\epsilon_1=0,
\end{align}
\begin{align}
\label{eq_epsilon2}
\epsilon_2 = -E_{\text{so}} ,
\end{align}
and
\begin{align}
\label{eq_epsilon34}
\epsilon_{3/4}=
\frac{1}{2} \left(-E_{\text{so}} \mp 
\sqrt{E_{\text{so}}^2+(2E_\Omega)^2} \right) .
\end{align}
The matrix representation $\underline{U}$
of $U$ and the basis states $| D_j \rangle$
are given in Appendix~\ref{appendix_rotation} for $\tilde{\delta}=0$.
For what follows, it is important that the basis states 
$| D_1 \rangle$, $| D_3 \rangle$, and $| D_4 \rangle$
are symmetric under the exchange of the two particles
while the basis state $| D_2 \rangle$
is
anti-symmetric under the exchange of the two particles.
Depending on the particle symmetry (BB versus FF), the 
combination of $z$- and $\rho$-dependent functions has to be chosen 
accordingly.

The approximate Hamiltonian $\hat{H}_{\text{rel}}^{UR,\text{sr}}$
[Eq.~(\ref{eq_hrel_ur})] 
is nearly identical to the full Hamiltonian $\hat{H}_{\text{rel}}$.
The difference is that the term proportional to
$\hat{p}_z$ in $\hat{H}_{\text{rel}}$ is replaced by 
the channel specific
energy shifts $\epsilon_j$ in $\hat{H}_{\text{rel}}^{UR,\text{sr}}$.
In addition to having gotten rid of the first derivative term,
the approximate Hamiltonian $\hat{H}_{\text{rel}}^{UR,\text{sr}}$
has another key characteristic: 
it is diagonal in the  $|D_j \rangle$ basis.
This implies that the propagation in the four channels labeled
by $|D_j \rangle$ can be done independently,
with the energy shift $\epsilon_j$ merely leading to a modified
scattering energy.
Thus, the approximate short-range Hamiltonian corresponds to
four ``standard'' wave guide problems in the absence of spin-orbit
coupling. Compared to the standard wave guide problem,
the scattering energy is replaced by the effective
scattering energies $E^{\text{eff}}_j=E- \epsilon_j$.
The propagation of these standard wave guide sub-systems
can be accomplished using essentially any propagator.

Having separately
propagated the four sub-systems of size
$n_{\text{max}} \times n_{\text{max}}$, 
the
logarithmic derivative
matrix 
$\underline{\cal{Y}}^{UR,\text{sr}}(z)$
of size $4n_{\text{max}} \times 4n_{\text{max}}$ 
is constructed
by combining the four $n_{\text{max}} \times n_{\text{max}}$
logarithmic derivative matrices. As in the full coupled-channel 
treatment, the channels are organized such that 
the four $n_{\rho}=0$ states are first, followed by the four $n_{\rho}=1$
states, and so on.
To transform from the 
$|D_j \rangle$ basis to the 
singlet-triplet basis, the rotation needs to be ``undone''.
The resulting expression for the
logarithmic derivative
matrix in the singlet-triplet basis reads
\begin{align}
\label{eq_ysoc}
\underline{\cal{Y}}^{\text{sr}}(z) =&
\left(\frac{d \underline{{R}}(z)}{dz} \right) 
\underline{{U}} \,
\left(\underline{{R}}(z) \, 
\underline{{U}} \right)^{-1}+ \nonumber \\
&\underline{{R}}(z) \,
\underline{{U}} \,  \underline{\cal{Y}}^{UR,\text{sr}}(z) \,
\left(\underline{{R}}(z) \, 
\underline{{U}} \right)^{-1}.
\end{align}

Equations~(\ref{eq_ysoc1}) and (\ref{eq_ysoc}) show that there are two terms
that contribute to the logarithmic derivative matrix
$\underline{\cal{Y}}^{\text{sr}}(z)$, which is
expressed in  the spin-orbit coupling basis
(i.e., in the same asymptotic basis as that used in
Sec.~\ref{sec_scatteringframework}). The first term 
on the right hand side of Eqs.~(\ref{eq_ysoc1})  and (\ref{eq_ysoc}) 
is due to the fact that the rotation operator
$\hat{R}$ is $z$-dependent. 
The matrix $\underline{\cal{Y}}^{\text{sr}}(z)$ couples,
just as the exact logarithmic derivative matrix $\underline{\cal{Y}}(z)$, 
different states of the
spin singlet-triplet basis.
Having $\underline{\cal{Y}}^{\text{sr}}(z)$,
the K-matrix is obtained by replacing
$\underline{\cal{Y}}(z)$ in Eq.~(\ref{eq_logderivative})
by $\underline{\cal{Y}}^{\text{sr}}(z)$.
We refer to the resulting K-matrix as
$\underline{K}^{\text{sr}}(z)$.
The superscript ``sr'' serves to remind the reader that
the K-matrix is obtained using the approximate 
small-$|z|$ logarithmic derivative matrix.

Importantly, the matching to the asymptotic solution
in the rotation approach is done in exactly the same way
as in the full coupled-channel treatment. This implies that
the procedure for determining and interpreting 
the physical K-matrix $\underline{K}_{\text{phys}}^{\text{sr}}$
obtained from $\underline{K}^{\text{sr}}$ is identical
to that outlined in Sec.~\ref{sec_physical}.
The ``only'' approximation made in the rotation approach is 
how the small-$|z|$ phase is being accumulated.

\begin{figure}[h]
\begin{center}
\includegraphics[width=2.6in]{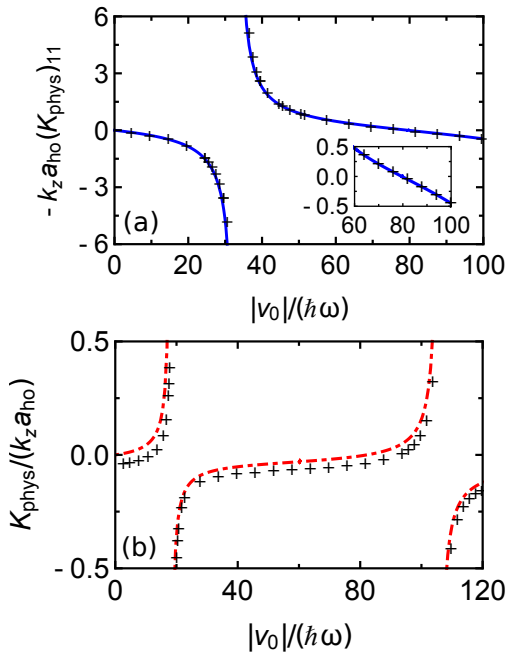}
\caption{(color online)
Benchmarking the rotation approach.
The lines and pluses
show scattering results obtained from the full propagation and the rotation approach,
respectively, for two identical fermions interacting through a Gaussian
potential with range $r_0=0.3a_{\text{ho}} / \sqrt{2}$ in the singlet channel
only 
as a function of the magnitude of the depth $v_0$ of the potential.
The results are obtained for $(k_{\text{so}})^{-1}=(0.2\sqrt{2})^{-1} a_{\text{ho}}$, 
$\tilde{\delta}=0$, and $E=E_{\text{th}}$.
Panel (a) shows the quantity 
$-k_z a_{\text{ho}}(\underline{K}_{\text{phys}}(k_z))_{11}$ 
for $E_\Omega = \hbar\omega/100$. 
The inset in (a) shows a blow-up of the main panel.
Panel (b) shows 
the quantity $K_{\text{phys}}(k_z)/(k_z a_{\text{ho}})$ for $E_\Omega=\hbar\omega$. 
}
\label{fig_rotation_singlet}
\end{center}
\end{figure}

\begin{figure}[h]
\begin{center}
\includegraphics[width=2.6in]{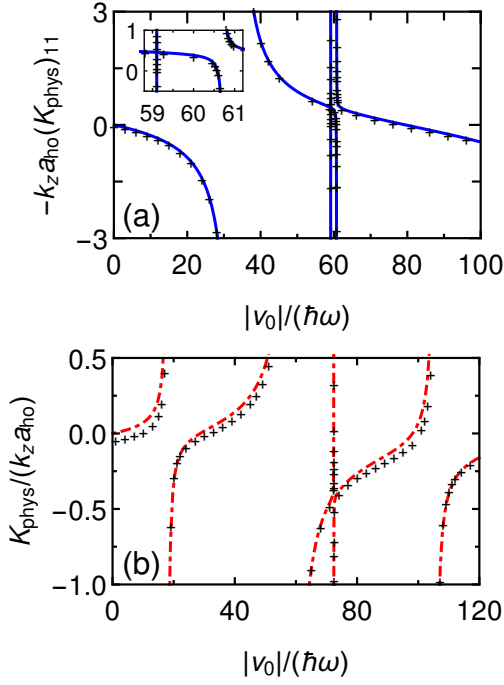}
\caption{(color online)
Benchmarking the rotation approach.
The lines and pluses
show scattering results obtained from the full propagation and the rotation approach,
respectively, for two identical fermions interacting through a Gaussian
potentials with range $r_0=0.3a_{\text{ho}} / \sqrt{2}$ in all four channels
as a function of the magnitude of the potential depth $v_0$.
The results are obtained for $(k_{\text{so}})^{-1}=(0.2\sqrt{2})^{-1} a_{\text{ho}}$, 
$\tilde{\delta}=0$, and $E=E_{\text{th}}$.
Panel (a) shows the quantity 
$-k_z a_{\text{ho}}(\underline{K}_{\text{phys}}(k_z))_{11}$ 
for $E_\Omega = \hbar\omega/100$. 
The inset in (a) shows a blow-up of the main panel.
Despite the enlarged scale, the resonance near $|v_0|=59 \hbar \omega$ is
not fully resolved.
Panel (b) shows 
the quantity $K_{\text{phys}}(k_z)/(k_z a_{\text{ho}})$ for $E_\Omega=\hbar\omega$. 
}
\label{fig_rotation_all}
\end{center}
\end{figure}

To assess the validity of the rotation approach, Fig.~\ref{fig_rotation_singlet}
compares scaled elements of the physical K-matrix
for two identical 
fermions with interaction in the singlet channel 
only obtained using the rotation approach (pluses)
and the full Hamiltonian (lines).
Figure~\ref{fig_rotation_all} considers two identical
fermions with identical interactions in all four channels.
In both figures, the 
non-vanishing interactions are modeled by a Gaussian potential
$V_{\text{G}}(r)$,
\begin{align}
\label{eq_gaussian_twobody}
  V_{\text{G}}(r) = v_0 \exp \left( - \frac{r^2}{2 r_0^2} \right),
  \end{align}
with range $r_0=0.3a_{\text{ho}}/ \sqrt{2}$ and varying
depth $v_0$ ($v_0 < 0$).
The scattering energy $E$ is set to $E_{\text{th}}$
and the generalized detuning 
$\tilde{\delta}$ to zero.

Figures~\ref{fig_rotation_singlet}(a) and \ref{fig_rotation_all}(a) 
show the quantity
$- k_z a_{\text{ho}} (\underline{K}_{\text{phys}}(k_z))_{11}$
(the reasoning behind this scaling is discussed in Sec.~\ref{sec_1dcoupling_soc})
for $E_{\Omega} = \hbar \omega/100$
(double-minimum regime) as a function of the magnitude of $v_0$. 
The agreement between the pluses and the lines indicates that
the rotation approach provides a quantitatively correct description of
the scattering observables in the small $E_{\Omega}$ regime.
Figures~\ref{fig_rotation_singlet}(b) and \ref{fig_rotation_all}(b) 
show the quantity
$K_{\text{phys}}(k_z)/(k_z a_{\text{ho}})$
for a larger $E_{\Omega}$
(namely, $E_{\Omega} = \hbar \omega$; single-minimum regime).
It can be seen that the results obtained by the rotation
approach deviate visibly from the full calculation.
Importantly, however, the rotation approach provides a semi-quantitatively
correct description even for this large Raman coupling strength.
In particular, the rotation approach reproduces the sharp resonance
feature around $|v_0|=70 \hbar \omega$ in Fig.~\ref{fig_rotation_all}(b).
Compared to the system with interaction in the singlet channel only, the
system with equal interactions in all four channels supports a richer
resonance structure. Our analysis shows that the resonances
near $|v_0|= 60 \hbar \omega$ in Fig.~\ref{fig_rotation_all}(a)
and
near $|v_0|= 60 \hbar \omega$
and $|v_0|= 70 \hbar \omega$ in Fig.~\ref{fig_rotation_all}(b)
involve the $p$-wave scattering volume.

\section{Effective one-dimensional coupling constants}
\label{sec_1dcoupling}

An alternative approach to determining the scattering solutions consists
of calculating effective
one-dimensional coupling constants
by ``integrating out'' the $n_{\rho}>0$ channels. The effective
one-dimensional coupling constants, in turn, provide the
input for an effective strictly one-dimensional 
$4 \times 4$ Hamiltonian $H_{\text{1d}}$. 
Assuming interactions in the singlet channel only, this approach
was pursued in Refs.~\cite{scientific_report,PRA_Zhang}.
To set the stage, Sec.~\ref{sec_waveguide}
reviews selected properties of the effective one-dimensional coupling constants
in the absence of spin-orbit coupling~\cite{Olshanii1998,GrangerBlume}. 
These reference results
will be very useful for interpreting the 
results in the presence of spin-orbit coupling.
Section~\ref{sec_1dcoupling_soc} discusses the effective one-dimensional
$4 \times 4$ Hamiltonian and analyzes the threshold behavior
in the vicinity of the lowest and higher-lying scattering thresholds.

\subsection{Reference system: Wave guide without spin-orbit coupling terms} 
\label{sec_waveguide}
In the absence of spin-orbit coupling 
($k_{\text{so}}=\Omega=\tilde{\delta}=0$), the four spin channels are 
decoupled and the problem reduces
to that of two particles in a wave guide. In this case, the
singlet channel is combined with an even-$z$
spatial wave function. In the limit of zero-range
interactions and a scattering energy
$E$ of $\hbar \omega$, 
this is the wave guide system considered in
Olshanii's seminal work~\cite{Olshanii1998}. 
Non-threshold scattering energies were subsequently considered in
Refs.~\cite{GrangerBlume,Panos_multichannel,CIREn,OlshaniiSchool}.
Each of the triplet channels is combined with an
odd-$z$ spatial wave function. In the limit
of short-range interactions, this is the wave guide system
considered by Granger and Blume~\cite{GrangerBlume}.
In the presence of the spin-orbit coupling terms, neither the
total spin nor the corresponding projection quantum number are conserved.
As a consequence, the singlet and triplet channels are mixed and the 
structure of the scattering resonances may be modified
compared to the scenarios without spin-orbit coupling terms.

We start our discussion with the singlet channel.
In what follows, we set the quantum number $m_l$ 
equal to $0$.
Modeling the interaction in the singlet
channel by a zero-range pseudo-potential characterized by
the free-space
$s$-wave scattering length $a_s$
and
enforcing that the spatial wave function is even with respect to
$z$, the effective one-dimensional interaction potential 
$V_{\text{1d}}^{\text{even}}(z)$
can be parametrized in terms of the one-dimensional
coupling constant $g_{\text{1d}}^{\text{even}}(k_{z})$~\cite{Olshanii1998,GrangerBlume},
\begin{eqnarray}
V_{\text{1d}}^{\text{even}}(z)= 
g_{\text{1d}}^{\text{even}}(k_{z}) \delta(z),
\end{eqnarray}
where $g_{\text{1d}}^{\text{even}}(k_{z})$ is given by
\begin{align}
\label{eq_g1deven}
\frac{g_{\text{1d}}^{\text{even}}(k_{z})}{\hbar \omega a_{\text{ho}}} = 
\frac{2 a_s(E)}{a_{\text{ho}}} \left[ 1 + \frac{a_s(E)}{a_{\text{ho}}}
\zeta \left(\frac{1}{2},\frac{3}{2} - \frac{E}{2 \hbar \omega} 
\right) \right]
^{-1},
\end{align}
$k_z$ denotes the
scattering wave number along $z$,
\begin{align}
E = \frac{\hbar^2 k_z^2}{2 \mu} + (2 n_{\rho} +1) \hbar \omega,
\end{align}  
$a_s(E)$ denotes the energy-dependent free-space $s$-wave scattering length,
$\zeta(\cdot,\cdot)$ denotes the Hurwitz-Zeta function,
and
the harmonic oscillator length $a_{\text{ho}}$ is defined
in terms of the reduced mass $\mu$, 
\begin{align}
a_{\text{ho}}=\sqrt{\frac{\hbar}{\mu \omega}}.
\end{align}
In what follows, we assume for simplicity that the 
scattering energy $E$ is chosen such that the $n_{\rho}=0$
channel is open while all other $n_{\rho}$ channels are closed
($\hbar \omega \le E < 3 \hbar \omega$).
Under this assumption,
the effective coupling constant $g_{\text{1d}}^{\text{even}}(k_z)$
is related to the physical K-matrix $K^{\text{even}}_{\text{phys}}(k_z)$ 
(this is a $1 \times 1$ matrix) via
\begin{align}
\label{eq_g1deven_k1d}
g_{\text{1d}}^{\text{even}}(k_z) = 
-\frac{\hbar^2 k_z}{\mu} K^{\text{even}}_{\text{phys}}(k_z).
\end{align}
The K-matrix diverges at the critical scattering length
$a_s^{\text{cr}}(E)$,
where 
\begin{align}
\label{eq_ascrit}
\frac{a_s^{\text{cr}}(E)}{a_{\text{ho}}}=-
\left[
\zeta \left(
\frac{1}{2}, \frac{3}{2} - \frac{E}{2 \hbar \omega}
\right)
\right]^{-1}.
\end{align}
For $E=E_{\text{th}}=\hbar \omega$, Eq.~(\ref{eq_ascrit})
reduces to $a_s^{\text{cr}}(E_{\text{th}}) \approx 0.6848 a_{\text{ho}}$~\cite{Olshanii1998,GrangerBlume}.
The resonance occurs when the scattering energy is equal to
the energy of a ``virtual bound state'' that is supported
by the
closed channel ($n_{\rho}>0$) Hilbert space~\cite{Olshanii2}. 
Within the zero-range framework,
the energy of the virtual
bound state lies exactly
$2 \hbar \omega$ above the energy of the true bound
state~\cite{Olshanii2} (throughout this manuscript,
we use the convention that true bound states,
calculated using
the Hilbert space spanned by the full Hamiltonian,
 have an energy
below the lowest relative scattering threshold).

In the literature, the interaction in the triplet channel
has been modeled by a $p$-wave zero-range 
pseudo-potential~\cite{KBpwave,Pricoupenko,3DPP,Jiang_pwave,Cui_Ebodd,ppDeutsch}.
Denoting the energy-dependent
free-space $p$-wave scattering volume by
$V_p(E)$ and enforcing that the spatial wave function
is
odd with respect to $z$, the effective
one-dimensional interaction potential $V_{\text{1d}}^{\text{odd}}(z)$
can be parametrized in terms of the one-dimensional
coupling constant $g_{\text{1d}}^{\text{odd}}(k_z)$~\cite{KBpwave},
\begin{eqnarray}
V_{\text{1d}}^{\text{odd}}(z) = 
g_{\text{1d}}^{\text{odd}} (k_z)
\overleftarrow{\frac{\partial}{\partial z}} \delta(z) \overrightarrow{\frac{\partial }{\partial z}},
\end{eqnarray}
where 
the first derivative operator acts to the left 
and the second to the right.
The derivative operators are needed since the 
spatial wave function vanishes at $z=0$.
Note that other parameterizations of the one-dimensional pseudo-potential
exist (see, e.g., Refs.~\cite{PP1D0,PP1D1,PP1D2,GO}).
Working in the low-energy regime where only
the $n_{\rho}=0$ channel is energetically open, 
the one-dimensional coupling constant
$g_{\text{1d}}^{\text{odd}}(k_z)$~\cite{GrangerBlume},
\begin{align}
\label{eq_g1dodd}
\frac{g_{\text{1d}}^{\text{odd}}(k_z)}{\hbar \omega a_{\text{ho}}^3} = \nonumber 
\\
\frac{- 6 V_p(E)}{a_{\text{ho}}^3} 
\left[
1- \frac{12 V_p(E)}{a_{\text{ho}}^3} 
\zeta
\left(-\frac{1}{2},\frac{3}{2} - \frac{E}{2 \hbar \omega} \right)
\right]^{-1},
\end{align}
is related to the physical K-matrix 
$K_{\text{phys}}^{\text{odd}}(k_z)$
through~\cite{GrangerBlume}
\begin{eqnarray}
\label{eq_g1dodd_k1d}
g_{\text{1d}}^{\text{odd}}(k_z) = 
\frac{\hbar^2}{\mu k_z} K_{\text{phys}}^{\text{odd}}(k_z).
\end{eqnarray}
The physical K-matrix
diverges at the critical scattering volume
$V_p^{\text{cr}}(E)$,
\begin{align}
\frac{V_p^{\text{cr}}(E)}{a_{\text{ho}}^3} =
\left[ 12 \zeta
\left(-\frac{1}{2},\frac{3}{2} - \frac{E}{2 \hbar \omega} \right)
\right]^{-1}
,
\end{align}
which reduces to
$V_p^{\text{cr}}(E_{\text{th}}) \approx -0.4009 a_{\text{ho}}^3$
for $E=E_{\text{th}}=\hbar \omega$.
According to Ref.~\cite{Jiang_pwave}, the resonance 
occurs---as in the even-$z$ case---when the scattering
energy is equal to the energy
of a virtual bound state 
in the closed channel ($n_{\rho}>0$)
Hilbert space.
Since the energy of the virtual bound state
coincides with
the true bound state
energy at threshold~\cite{Jiang_pwave}, 
the emergence of the true bound state can be 
used to identify the resonance positions of
$K_{\text{phys}}^{\text{odd}}(k_z)$ for $k_z=0$.
Below the lowest scattering threshold, the energy of the virtual bound state 
deviates from the energy of the true bound state;
in particular, the energy of the 
virtual bound state is not, as in the even-$z$ case, shifted up
by a constant with respect to the energy of the true bound state.

\begin{figure}[h]
\begin{center}
\includegraphics[width=2.6in]{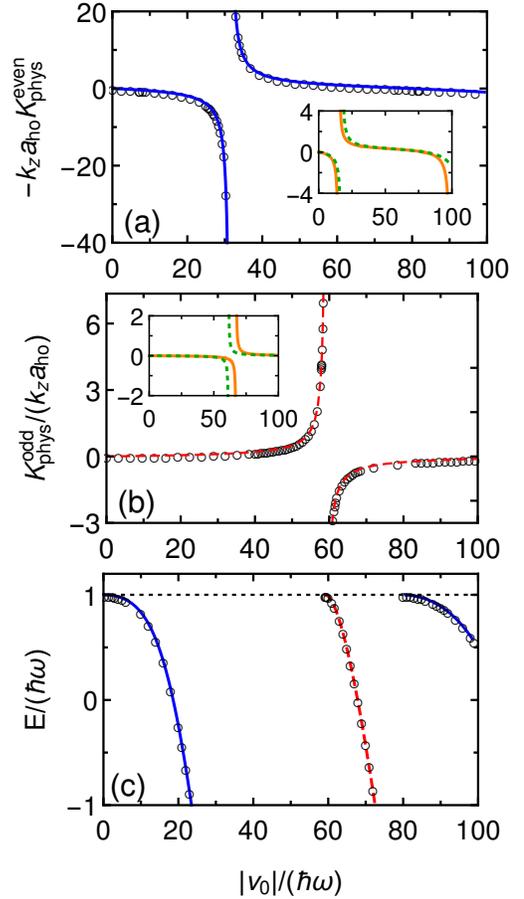}
\caption{(color online)
Wave guide properties in the absence of spin-orbit
coupling as a function of the magnitude of the depth
$v_0$ of the Gaussian potential with $r_0=0.3a_{\text{ho}} / \sqrt{2}$.
(a) $\lim_{k_z \rightarrow 0} 
[- k_z a_{\text{ho}} K_{\text{phys}}^{\text{even}}(k_z)]$.
The solid line shows finite-range
results. The open circles show zero-range results
that are obtained by taking the energy-dependence of the free-space
scattering quantities into account.
The solid and dotted lines in the inset of panel~(a)
show 
$a_s(0)/a_{\text{ho}}$ and $a_s(\hbar \omega)/a_{\text{ho}}$, 
respectively,
as a function of $|v_0|/(\hbar \omega)$.
(b) $\lim_{k_z \rightarrow 0}  
[K_{\text{phys}}^{\text{odd}}(k_z)/( k_z a_{\text{ho}})]$.
The dashed line shows finite-range
results. The open circles show zero-range results
that are obtained by taking the energy-dependence of the free-space
scattering quantities into account.
The solid and dotted lines in the inset of panel~(b)
show 
$V_p(0)/a_{\text{ho}}^3$ and $V_p(\hbar \omega)/a_{\text{ho}}^3$, 
respectively,
as a function of $|v_0|/(\hbar \omega)$.
(c) Relative energy of the even-$z$ (solid lines) and odd-$z$ 
(dashed line) bound states for the finite-range potential.
The open circles show the zero-range results.
To aid the readability, the horizontal dashed line indicates the 
threshold energy $E_{\text{th}}$.
}
\label{fig_nosoc}
\end{center}
\end{figure}

Our calculations in the presence of the spin-orbit coupling
terms are not performed for a zero-range interaction potential
but for the Gaussian interaction
potential $V_{\text{G}}(r)$ [see Eq.~(\ref{eq_gaussian_twobody})].
The solid line in Fig.~\ref{fig_nosoc}(a)
and the dashed line in Fig.~\ref{fig_nosoc}(b) show the
scaled physical K-matrices 
$-k_z a_{\text{ho}} K_{\text{phys}}^{\text{even}}(k_z)$
and $K_{\text{phys}}^{\text{odd}}(k_z)/ ( k_z a_{\text{ho}})$, 
respectively,
in the $k_z \rightarrow 0$ limit
as a function of
the magnitude of the depth $v_0$ of the
Gaussian interaction potential with range $r_0=0.3 a_{\text{ho}}/ \sqrt{2}$.
For comparison, the open
circles in Fig.~\ref{fig_nosoc}(a) show the zero-range
result for the scaled K-matrix 
$- k_z a_{\text{ho}} K_{\text{phys}}^{\text{even}}(k_z)$,
which is obtained by using Eqs.~(\ref{eq_g1deven})
and (\ref{eq_g1deven_k1d}) with $a_s(\hbar \omega)$;
this quantity coincides with 
$g_{\text{1d}}^{\text{even}} / (\hbar \omega a_{\text{ho}})$.
Similarly,
the open circles in Fig.~\ref{fig_nosoc}(b) show the zero-range
result for the scaled K-matrix $K_{\text{phys}}^{\text{odd}}(k_z)/( k_z a_{\text{ho}})$,
which is obtained by using Eqs.~(\ref{eq_g1dodd})
and (\ref{eq_g1dodd_k1d}) with $V_p(\hbar \omega)$;
this quantity coincides with 
$g_{\text{1d}}^{\text{odd}} / (\hbar \omega a_{\text{ho}}^3)$.
For the relatively
large range $r_0$ of the Gaussian interaction potential considered, 
the inclusion of the energy dependence of the $s$-wave scattering length
and $p$-wave scattering volume is crucial
[see insets of Figs.~\ref{fig_nosoc}(a) and 
\ref{fig_nosoc}(b)]~\cite{3DPP,Naidon}.

The solid and dashed lines in Fig.~\ref{fig_nosoc}(c) 
show the relative energy of, respectively, 
the even-$z$ and odd-$z$
bound states for the Gaussian potential.
The two-body system in a wave guide is bound if the 
relative energy is smaller than $\hbar \omega$.
Compared to the free-space
case, where the system is bound when the
relative energy is smaller than zero, the wave guide leads to an enhancement of
the binding of the most weakly-bound state. For example, the 
wave guide supports a weakly-bound even-$z$ state for $|v_0|$ greater
than zero while the system without a wave guide supports a weakly-bound 
$s$-wave state for $|v_0|$ greater than $14.91 \hbar \omega$.
Similarly,
the wave guide supports a weakly-bound odd-$z$ state for $|v_0|$ greater
than $59.24 \hbar \omega$
while the system without a wave guide supports a weakly-bound 
$p$-wave state for $|v_0|$ greater than $67.22 \hbar \omega$.

For comparison, the open circles in Fig.~\ref{fig_nosoc}(c)
show the zero-range
results, which are obtained by solving the 
zero-range eigen energy equations self-consistently,
accounting for the energy-dependence of
the $s$-wave scattering length
and $p$-wave scattering volume~\cite{footnote1}.
The agreement between the 
zero-range and finite-range binding energies 
is quite good on the scale shown, demonstrating that the zero-range treatment
provides a reliable description even though the range of
our two-body Gaussian potential is quite large.
We note, however, that the extremely small finite-range binding
energies 
near $|v_0| \approx 80 \hbar \omega$ 
and 
near $|v_0| \approx 60 \hbar \omega$ 
are not well reproduced by the zero-range
models. The reason is that the quantities 
$a_s/r_0$ and $|V_p|^{1/3}/r_0$,
respectively, are not notably smaller than one
for these $|v_0|$.

Figure~\ref{fig_nosoc} serves as a reference for the calculations
presented in Sec.~\ref{sec_results}, which account
for the
spin-orbit coupling terms.
In particular, that section explores how the resonances 
shift as a function of the spin-orbit coupling parameters.

\subsection{Effective one-dimensional $4 \times 4$ low-energy
Hamiltonian}
\label{sec_1dcoupling_soc}

Assuming $s$-wave zero-range interactions 
(modeled using the three-dimensional Fermi-Huang pseudo-potential)
in the singlet channel
only and setting $m_l$ to zero, 
Refs.~\cite{scientific_report,PRA_Zhang} derived the following
effective strictly
one-dimensional Hamiltonian
$H_{\text{1d}}$ in the relative coordinate $z$
for two identical fermions,
\begin{align}
\label{eq_effective_1d_ham}
H_{\text{1d}} = 
\left( -\frac{\hbar^2}{2 \mu} \frac{\partial^2 }{\partial z^2} 
+
\hbar \omega
\right)
\hat{I}_1 \otimes
\hat{I}_2 + 
\hat{V}_{\text{so}} + \nonumber \\
g_{\text{1d}}^{\text{soc}}(k_z) \delta(z) | S_0 \rangle \langle S_0| 
.
\end{align}
The quantity $\hat{V}_{\text{so}}$ is defined in Eq.~(\ref{eq_vso})
and the one-dimensional coupling constant
$g_{\text{1d}}^{\text{soc}}(k_z)$
results from ``integrating 
out'' the $n_{\rho}>0$ harmonic oscillator channels.
If the scattering energy $E$ is chosen such that only a subset of the
four $n_{\rho}=0$ channels is open, then $g_{\text{1d}}^{\text{soc}}$
corresponds to an ``unphysical effective coupling constant''.
The ``physical effective coupling constant'' is obtained
by eliminating the energetically closed
$n_{\rho}=0$ channels from $\underline{K}_{\text{1d}}$ (see below).
Importantly, $\underline{K}_{\text{1d}}$ is derived from a
three-dimensional Hamiltonian; the ``1d'' subscript is used to
distinguish the zero-range treatment (Refs.~\cite{scientific_report,PRA_Zhang})
from the finite-range
treatment discussed in the previous sections.
We emphasize that
the derivation of the effective one-dimensional Hamiltonian
given in Eq.~(\ref{eq_effective_1d_ham}) assumes that all
$n_{\rho}>0$ channels are energetically closed~\cite{scientific_report,PRA_Zhang}.
This restricts the scattering energy to
values $E < E_{\text{th}}+2 \hbar \omega$. 
For certain relatively large $E_{\Omega}$ or $E_{\text{so}}$, the 
scattering energy may be further restricted
due to a ``reordering'' of the non-interacting relative
dispersion curves (i.e., the minimum
of a dispersion curve with $n_{\rho}=0$ may
lie above the minimum of a dispersion curve with $n_{\rho}=1$).
This is discussed further
in Sec.~\ref{sec_singlet_triplet}.

\begin{figure}[h!]
\begin{center}
\includegraphics[width=2.8in]{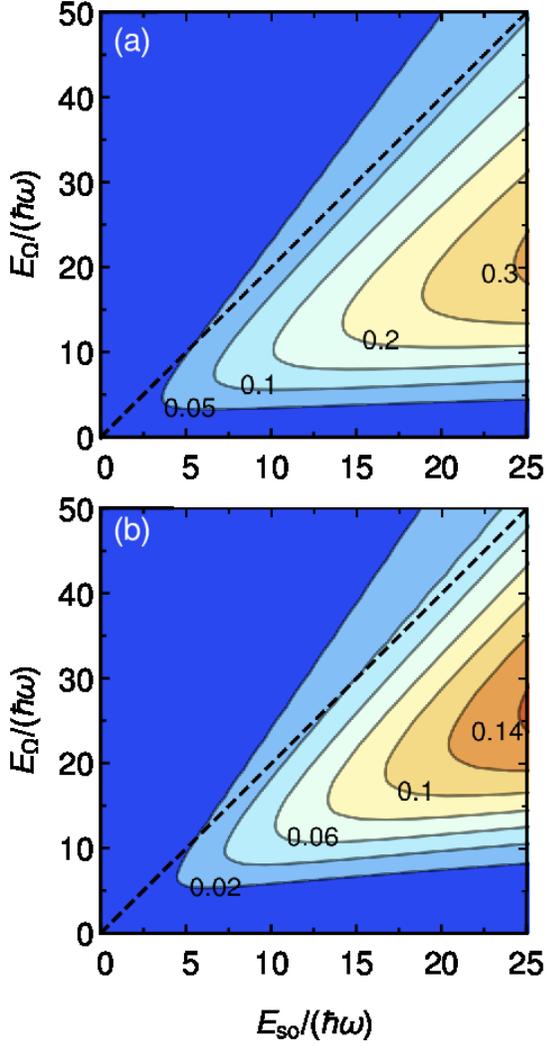}
  \caption{(color online)
Benchmarking the accuracy of the Taylor expansion
given in Eq.~(\ref{eq_capc_approx}), which provides a
simple approximate means to evaluate the effective one-dimensional
coupling constant $g_{\text{1d}}^{\text{soc}}$.
Panels~(a) and (b) show contour plots of the fractional differences
    $(C-D_0)/C$
and
$(C-D_0-D_1)/C$, respectively,
     as functions of $E_{\Omega}/(\hbar \omega)$
    and
    $E_{\text{so}}/(\hbar \omega)$
  for $E=E_{\text{th}}$.
The contours are spaced equidistantly (selected contours are labeled).
  The dashed lines mark the transition from
the double-minimum to the single-minimum regime.
    }
\label{fig_percentage}
\end{center}
\end{figure}

For $\tilde{\delta}=0$, Ref.~\cite{scientific_report} 
writes
$g_{\text{1d}}^{\text{soc}}$ as
(see the last two
equations before the reference section) 
\begin{align}
  \label{eq_g1dsoc}
  \frac{g_{\text{1d}}^{\text{soc}}}{\hbar \omega a_{\text{ho}}} =
  \frac{2 a_s}{a_{\text{ho}}} 
  \left[
    1 
+ \frac{a_s}{a_{\text{ho}}}    
C \left(\frac{E}{\hbar \omega},
\frac{E_{\Omega}}{\hbar \omega},
\frac{E_{\text{so}}}{\hbar \omega} \right) 
 \right] ^{-1}
  \end{align}
with
\begin{align}\label{eq_capc}
C \left(\frac{E}{\hbar \omega},
\frac{E_{\Omega}}{\hbar \omega},
\frac{E_{\text{so}}}{\hbar \omega} \right) =
-
C_1\left(\frac{E}{\hbar \omega}\right) 
-
C_2 \left(\frac{E}{\hbar \omega},
\frac{E_{\Omega}}{\hbar \omega},
\frac{E_{\text{so}}}{\hbar \omega} \right) ,
\end{align}
where the functions $C_1$ and $C_2$
are given in terms
of one- and two-dimensional integrals, respectively.
Equation~(\ref{eq_g1dsoc}) has the same functional form as Eq.~(\ref{eq_g1deven}), 
except that 
the Hurwitz-Zeta function is replaced by the three-parameter function $C$.
Taylor-expanding the function $C$ about 
$E_{\Omega}/(\hbar \omega)=0$,
we find 
\begin{align}
  \label{eq_capc_approx}
C \left(\frac{E}{\hbar \omega},
\frac{E_{\Omega}}{\hbar \omega},
\frac{E_{\text{so}}}{\hbar \omega} \right) \approx
D_0 \left( \frac{E+E_{\text{so}}}{\hbar \omega} \right) + \nonumber \\
D_1 \left(
\frac{E}{\hbar \omega},\frac{E_{\text{so}}}{\hbar \omega},\frac{E_{\Omega}}{\hbar \omega}
\right) + 
{\cal{O}}\left(
\frac{E_{\Omega}^4}{(\hbar \omega)^4}
\right),
\end{align}
where
\begin{align}
\label{eq_capcbar0}
D_0 \left( \frac{E+E_{\text{so}}}{\hbar \omega} \right)=
\zeta \left( \frac{1}{2} , \frac{3}{2} - \frac{(E+E_{\text{so}})}{2 \hbar \omega}
\right)
\end{align}
and
\begin{align}
\label{eq_capcbar2}
D_1 \left(
\frac{E}{\hbar \omega},\frac{E_{\text{so}}}{\hbar \omega},\frac{E_{\Omega}}{\hbar \omega}
\right)
=- 
\frac{E_{\Omega}^2}{(\hbar \omega)^2} \left(\frac{E_{\text{so}}}{\hbar \omega} \right)^{-1}
\times \nonumber \\
\Bigg[
\frac{1}{8} \zeta \left( \frac{3}{2},\frac{3}{2}-\frac{E}{2\hbar \omega} \right) -                                                    
\frac{1}{8} \zeta \left( \frac{3}{2}, \frac{3}{2}-\frac{E + E_{\text{so}}}{2 \hbar \omega} \right) +\nonumber \\     
\sum_{n=0}^{\infty} 
\frac{1}{4 \sqrt{\pi} n! (2n+1)} \left( \frac{E_{\text{so}}}{2 \hbar \omega} \right)^{n+1}
\Gamma \left( n+\frac{5}{2} \right) \times \nonumber \\
\zeta \left(
n+\frac{5}{2},\frac{3}{2} -\frac{E}{2 \hbar \omega}   
\right)
\Bigg].
\end{align}
The sum over $n$ on the right hand side of Eq.~(\ref{eq_capcbar2}) converges 
relatively quickly.
Figures~\ref{fig_percentage}(a) and \ref{fig_percentage}(b) show the fractional differences
$(C-D_0)/C$ and $(C-D_0-D_1)/C$, respectively,
for $E=E_{\text{th}}$. It can be seen that the fractional difference
$(C-D_0)/C$ is smaller than about 30\%
for the ranges of $E_{\Omega}/(\hbar \omega)$ 
and $E_{\text{so}}/(\hbar \omega)$ considered.
Moreover, Fig.~\ref{fig_percentage}(b) demonstrates that the approximation
provides a description at an accuracy of 2\% or better 
for a wide range of parameter combinations.
This suggests that the approximate but compact 
expansion can be used over a fairly large parameter
space unless high accuracy
results are sought.

The Taylor expansion given in Eq.~(\ref{eq_capc_approx})
can be understood from the analysis presented in Sec.~\ref{sec_rotation}.
To see this we consider the small-$|z|$
Hamiltonian $\hat{H}_{\text{rel}}^{UR,\text{sr}}$ [Eq.~(\ref{eq_hrel_ur})],
which is diagonal in the $|D_j \rangle $ basis.
The derivation of  $\hat{H}_{\text{rel}}^{UR,\text{sr}}$ assumed
that the interactions in all four channels are equal.
It may thus seem that we cannot apply it to the case with
interaction in the singlet channel only considered here.
However, if we add the Fermi-Huang pseudo-potential
to the three triplet channels, the zero-range results
summarized above remain unaltered since the Fermi-Huang pseudo-potential only
acts when the spatial wave function component is non-zero at $\vec{r}=0$.
This implies that the Hamiltonian $\hat{H}_{\text{rel}}^{UR,\text{sr}}$
applies to the case considered here.
Combining the energy shift  $\epsilon_2=-E_{\text{so}}$
in the anti-symmetric $|D_2 \rangle $ channel with the scattering energy
$E$ [see Eqs.~(\ref{eq_hrel_ur}) and (\ref{eq_epsilon2})],
the phase shift that is being accumulated in the
$|D_2 \rangle$ channel is the same as the one that
would be accumulated in the absence of the spin-orbit
coupling terms for the energy $E + E_{\text{so}}$ as opposed to
for $E$. This explains why $g_{\text{1d}}^{\text{soc}}$ in
Eq.~(\ref{eq_g1dsoc})
has, approximately, the same functional form
as in the absence of spin-orbit coupling, except
that the scattering energy is
replaced by the effective scattering energy $E + E_{\text{so}}$.
Correspondingly, the higher-order corrections in the 
Taylor expansion in Eq.~(\ref{eq_capc_approx}) can
be interpreted as representing the
correction terms that would arise in the rotation approach
if higher-order terms in $k_{\text{so}} |z|$ were taken into account
in Eq.~(\ref{eq_hrel_ur}).

When Eq.~(\ref{eq_g1dsoc}) is applied to predict or to reproduce results
for finite-range potentials,
the energy-dependence of $a_s$ should be taken into account.
Section~\ref{sec_results} evaluates the scattering length $a_s$
that enters into
$g_{\text{1d}}^{\text{soc}}(k_z)$ [see Eq.~(\ref{eq_g1dsoc})]
at the 
scattering energy $E$.

Section~\ref{sec_results}
compares the physical K-matrix $\underline{K}_{\text{phys}}$ for
finite-range interactions,
obtained using the formalism discussed in Sec.~\ref{sec_scatteringframework},
with the physical K-matrix $\underline{K}_{\text{1d,phys}}$ obtained by
determining the scattering solutions for the
effective one-dimensional Hamiltonian.
In what follows, we restrict our discussion to
the $\tilde{\delta}=0$ case.
The $4 \times 4$ K-matrix $\underline{K}_{\text{1d}}$
for the strictly one-dimensional
Hamiltonian can be readily obtained analytically
in terms of $g_{\text{1d}}^{\text{soc}}(k_z)$, $E$, and the spin-orbit
coupling parameters.
Due to their lengthy-ness, the expressions are not reproduced here.
Eliminating the energetically closed channels,
the physical K-matrix $\underline{K}_{\text{1d,phys}}$
can be obtained for the double-minimum and single-minimum regimes.

To illustrate the impact of the modified single-particle dispersion
curves on the scattering properties, 
we analyze the behavior of $\underline{K}_{\text{1d,phys}}$
in the limit that the scattering energy $E$ approaches the
energy $E_{\alpha}$, where $\alpha$ is equal
to $I$, $II$, $II'$, $III$, or $IV$
(see the caption of Fig.~\ref{fig1} for the definition of
$E_{\alpha}$).
For $\alpha=I$ and $II$
($E_I$ and $E_{II}$ are the threshold energies
in the double-minimum and single-minimum regimes, 
respectively), we consider the
$E \rightarrow (E_{\alpha})^{+}$ limit.
For the other $\alpha$, we consider the
$E \rightarrow (E_{\alpha})^{+}$ 
and
$E \rightarrow (E_{\alpha})^{-}$ 
limits.
Using the quantities ${\bar{k}}_{\alpha}^{(\pm)}$,
\begin{align}
{\bar{k}}_{\alpha}^{(\pm)} = \left(\pm \frac{m}{\hbar^2} (E- E_{\alpha}) \right)^{1/2}, 
\end{align}
as small parameters,
the near-threshold behavior 
of the physical K-matrix $\underline{K}_{\text{1d,phys}}$
and its eigenvalues $K_{\text{1d,phys}}^{(j)}$ 
is summarized in Table~\ref{tab_threshold}.
Note that the ${\bar{k}}_{\alpha}^{(\pm)}$ are defined as real, positive quantities,
with the superscripts ``$(+)$'' and ``$(-)$''
referring to the 
$E \rightarrow (E_{\alpha})^{+}$ 
and
$E \rightarrow (E_{\alpha})^{-}$ 
limits.
The coefficients $a_{\alpha,s}^{(\pm)}$,
$b_{\alpha,s}^{(\pm)}$, and $c_{\alpha,s}^{(\pm)}$
are determined by 
$g_{\text{1d}}^{\text{soc}}$,
$a_{\text{ho}}$, $\hbar \omega$, $E_\Omega$, and $E_{\text{so}}$.
In general, the analytic expressions for
$a_{\alpha,s}^{(\pm)}$,
$b_{\alpha,s}^{(\pm)}$, and $c_{\alpha,s}^{(\pm)}$ are quite involved.
As examples, we consider
$a_{I,-1}^{(+)}$ and $a_{II,1}^{(+)}$,
which can both be written in the form
(the functional form for the other coefficients may be different)
\begin{align}
\label{eq_coeff_threshold}
a_{\alpha,s}^{(\pm)}=
A_{\alpha,s}^{(\pm)} \frac{g_{\text{1d}}^{\text{soc}}}{a_{\text{ho}} \hbar \omega} 
\left( 
1+ \frac{g_{\text{1d}}^{\text{soc}}}{a_{\text{ho}} \hbar \omega}
B_{\alpha,s}^{(\pm)} 
\right)^{-1}
\end{align}
with
\begin{align}
A_{I,-1}^{(+)}=
-
\frac{\left[(2 E_{\text{so}})^2 - E_{\Omega}^2 \right]^{1/2}}{4 a_{\text{ho}} E_{\text{so}}} 
,
\end{align}
\begin{align}
B_{I,-1}^{(+)}=
-
\frac{E_{\Omega}^2 (\hbar \omega)^{1/2}}
{(2E_{\text{so}})^{3/2}  \left[(2 E_{\text{so}})^2 + E_{\Omega}^2 \right]^{1/2}}
,
\end{align}
\begin{align}
\label{eq_aroman21_plus}
A_{II,1}^{(+)}=
\frac{a_{\text{ho}}
\hbar\omega}{ \left(E_{\Omega}^2-2E_{\text{so}} E_\Omega\right)^{1/2}}  
\frac{E_{\text{so}}(4E_{\text{so}}-E_\Omega)}
{(E_\Omega^2-6E_{\text{so}}E_\Omega+8E_{\text{so}}^2)},
\end{align}
and
\begin{align}
\label{eq_broman21_plus}
B_{II,1}^{(+)}= \nonumber \\
\frac{\sqrt{\hbar\omega}
\left[ E_\Omega^{3/2}-
2E_{\text{so}}E_\Omega^{1/2}-
2\sqrt{2}E_{\text{so}}
\left( E_\Omega-2E_{\text{so}}\right)^{1/2} \right]}{\sqrt{2}(E_\Omega^2-6E_{\text{so}}E_\Omega+8E_{\text{so}}^2)}.
\end{align}
The coefficients $-2\hbar^2a_{I,-1}^{(+)}/\mu$ and $\hbar^2a_{II,1}^{(+)}/\mu$ can be
interpreted as effective
even- and odd-$z$ coupling constants. This interpretation is motivated by
the fact that the associated non-zero eigen values $K_{\text{1d,phys}}^{(1)}$
(see Table~\ref{tab_threshold})
scale in the same way 
as Eqs.~(\ref{eq_g1dodd_k1d}) and (\ref{eq_g1deven_k1d}),
respectively.
While not
pursued in this work,
the effective even-$z$ coupling constant
$-2\hbar^2 a_{I,-1}^{(+)}/\mu$ and the effective odd-$z$ coupling
constant $\hbar^2 a_{II,1}^{(+)}/\mu$ should provide the starting point for
developing an effective low-energy single-band description
of the system.
We note that the threshold behavior and structure 
of the coefficients
can also, with quite a bit of work, be deduced from the 
results presented in the {\em{Method Section}} of 
Ref.~\cite{scientific_report}.

\begin{widetext}

  \begin{table}
  \begin{center}
  \begin{tabular}{l|c|c}
 & $\underline{K}_{\text{1d,phys}}$ & $K_{\text{1d,phys}}^{(j)}$ \\ \hline
$E \rightarrow (E_I)^+$ & $\left( \begin{array}{cc}
                           a_{I,-1}^{(+)} ({\bar{k}}_{I}^{(+)})^{-1} + a_{I,0}^{(+)} 
&
                          - a_{I,-1}^{(+)} ({\bar{k}}_{I}^{(+)})^{-1}  
\\
                          - a_{I,-1}^{(+)} ({\bar{k}}_{I}^{(+)})^{-1}  
&                            a_{I,-1}^{(+)} ({\bar{k}}_{I}^{(+)})^{-1} - a_{I,0}^{(+)} 
\end{array} \right)$ & 
$\begin{array}{l}
 j=1: 2 a_{I,-1}^{(+)} ({\bar{k}}_{I}^{(+)})^{-1} \\
j=2: 0 \end{array}$ \\ \hline
$E \rightarrow (E_{II})^+$ & 
$a_{II,1}^{(+)}{\bar{k}}_{II}^{(+) } + a_{II,3}^{(+)}({\bar{k}}_{II}^{(+) })^3 $ & 
\\ \hline
$E \rightarrow (E_{II'})^+$ & 
$a_{II',0} + a_{II',1}^{(+)} {\bar{k}}_{II'}^{(+)}$ & 
\\ \hline
$E \rightarrow (E_{II'})^-$ & 
$\left( \begin{array}{cc}
                           a_{II',0}  &
                           a_{II',1/2}^{(-)} ({\bar{k}}_{II'}^{(-)})^{1/2}  \\
                           a_{II',1/2}^{(-)} ({\bar{k}}_{II'}^{(-)})^{1/2}   &
                           a_{II',1}^{(-)} {\bar{k}}_{II'}^{(-)} 
\end{array} \right)$  & 
$\begin{array}{l}
j=1:a_{II',0} + a_{II',1}^{(-)} {\bar{k}}_{II'}^{(-)} 
\\
j=2: 0 \end{array}$
\\ \hline
$E \rightarrow (E_{III})^+$ &   
$\left( \begin{array}{ccc}
                           a_{III,0}^{(+)}   &
                           0 & 
                           a_{III,-1/2}^{(+)} ({\bar{k}}_{III}^{(+)})^{-1/2}\\
                           0&0&0\\
                           a_{III,-1/2}^{(+)} ({\bar{k}}_{III}^{(+)})^{-1/2}   &                           
                           0&
                           a_{III,-1}^{(+)} ({\bar{k}}_{III}^{(+)})^{-1}+ a_{III,1}^{(+)} {\bar{k}}_{III}^{(+)}
\end{array} \right)$  &
$\begin{array}{l}
 j=1: a_{III,-1}^{(+)}({\bar{k}}_{III}^{(+)})^{-1}+ a_{III,0}^{(+)} \\
j=2: 0 \\
j=3: 0\end{array}$
 \\ \hline
$E \rightarrow (E_{III})^-$ & 
 $a_{III,1}^{(-)} {\bar{k}}_{III}^{(-) }+ a_{III,2}^{(-)}  ({\bar{k}}_{III}^{(-)})^{2} $ &
  \\ \hline
$E \rightarrow (E_{IV})^+$ & 
$\left( \begin{array}{cccc}
                           a_{IV,0}  &
                           0&
                           b_{IV,0}^{(+)}& 
                           a_{IV,1/2}^{(+)} ({\bar{k}}_{IV}^{(+)})^{1/2}\\
                           0&0&0&0\\                           
                           b_{IV,0}^{(+)} &                           
                           0&
                           c_{IV,0}&
                           b_{IV,1/2}^{(+)} ({\bar{k}}_{IV}^{(+)})^{1/2}\\
                           a_{IV,1/2}^{(+)} ({\bar{k}}_{IV}^{(+)})^{1/2}&
                           0&
                           b_{IV,1/2}^{(+)} ({\bar{k}}_{IV}^{(+)})^{1/2}&
                           a_{IV,1}^{(+)} {\bar{k}}_{IV}^{(+)}
\end{array} \right)$&
$\begin{array}{l}
j=1: (a_{IV,0}+c_{IV,0})+ a_{IV,1}^{(+)} {\bar{k}}_{IV}^{(+)} \\
j=2: 0 \\
j=3: 0\\
j=4:0 \end{array}$
 \\ \hline
$E \rightarrow (E_{IV})^-$ &
$\left( \begin{array}{ccc}
                           a_{IV,0}+a_{IV,1}^{(-)} {\bar{k}}_{IV}^{(-)} &
                           0&
                           b_{IV,0}^{(-)}+b_{IV,1}^{(-)} {\bar{k}}_{IV}^{(-)} \\
                           0&0&0\\       
                           b_{IV,0}^{(-)}+b_{IV,1}^{(-)} {\bar{k}}_{IV}^{(-)} &  
                           0&                         
                           c_{IV,0}+c_{IV,1}^{(-)} {\bar{k}}_{IV}^{(-)}
\end{array} \right)$& 
$\begin{array}{l}
j=1: (a_{IV,0}+c_{IV,0})+(a_{IV,1}^{(-)}+c_{IV,1}^{(-)})  {\bar{k}}_{IV}^{(-)} \\
j=2: 0 \\
j=3: 0\end{array}$
\end{tabular}
\caption{Summary of threshold laws for two identical fermions
with $\tilde{\delta}=0$
obtained by analyzing the zero-range model with interaction in the singlet channel
only.
In the cases where $\underline{K}_{\text{1d,phys}}$ is a 
$1 \times 1$ matrix, the eigen value $K_{\text{1d,phys}}^{(1)}$ is not retyped in the
$K_{\text{1d,phys}}^{(j)}$ column.
The energies $E_{\alpha}$, where $\alpha$ takes the Roman 
letters $I$, $II$, $II'$, $III$, and $IV$, are defined in the caption
of Fig.~\protect\ref{fig1}.
The quantities $a_{\alpha,s}^{(\pm)}$, $b_{\alpha,s}^{(\pm)}$,
and $c_{\alpha,s}^{(\pm)}$
 are constants that
depend on
$E_{\Omega}$, $E_{\text{so}}$, $\hbar \omega$, $a_{\text{ho}}$,
and $g_{\text{1d}}^{\text{soc}}$
(they are independent of $E$). The subscript
$s$ indicates the power of ${\bar{k}}_{\alpha}^{(\pm)}$ that the coefficient is
associated with.
Note the following identities: $a_{II',0}^{(+)}=a_{II',0}^{(-)}=a_{II',0}$,
$a_{IV,0}^{(+)} =a_{IV,0}^{(-)}=a_{IV,0}$, and
$c_{IV,0}^{(+)} =c_{IV,0}^{(-)}=c_{IV,0}$.
The eigenvalues $K_{\text{1d,phys}}^{(j)}$ with $j>1$ 
vanish identically for all energies, not just in the vicinity of
the thresholds.
}
\label{tab_threshold}
\end{center}
\end{table}

\end{widetext}

Inspection of Table~\ref{tab_threshold} shows that
the eigen value
$K_{\text{1d,phys}}^{(1)}$ has the characteristics
of the usual odd-$z$ threshold behavior
in the vicinity of $(E_{II})^{+}$ and $(E_{III})^{-}$ and 
the usual even-$z$ threshold behavior
in the vicinity of $(E_{I})^+$ and $(E_{III})^{+}$. 
In the vicinity of $(E_{II'})^{\pm}$ and $(E_{IV})^{\pm}$, the 
threshold behavior of the eigen value
$K_{\text{1d,phys}}^{(1)}$
is unusual (no energy dependence). This demonstrates
the non-trivial impact of the spin-orbit coupling terms
in the low-energy regime.
We note that 
some of the matrix elements of the physical K-matrix near the scattering
thresholds contain half-integer powers of the small wave vector $\bar{k}_{\alpha}^{(\pm)}$.
Since the scaling with $\bar{k}_{\alpha}^{(\pm)}$
changes across some of the higher-lying scattering thresholds,
Table~\ref{tab_threshold} suggests
that the scattering observables across some scattering thresholds
may not be smooth. 
This will be elaborated on further in
Sec.~\ref{sec_singlet_triplet}.
A ``jump'' in the total reflection
coefficient across one of the scattering thresholds was already pointed out in
the context of Fig.~2 of Ref.~\cite{scientific_report}.

\section{Scattering observables for two identical fermions with
interaction in the singlet channel}
\label{sec_results}
This section considers the effect of the spin-orbit coupling terms 
on the scattering observables 
for two identical fermions with interaction in the singlet channel
only for
$\tilde{\delta}=0$
and a $(k_{\text{so}})^{-1}$ 
that is much larger than
the transverse confinement length $a_{\text{ho}}$,
namely $(k_{\text{so}})^{-1} \approx 3.54 a_{\text{ho}}$.
Various 
Raman coupling strengths $\Omega$ and potential depths $v_0$
are considered.
A naive expectation might be 
that a small $k_{\text{so}}$ can only weakly
perturb the scattering properties obtained 
in the absence of spin-orbit coupling.
This section shows
that this naive expectation is not necessarily
correct, i.e., the spin-orbit coupling terms can induce
significant changes even for $(k_{\text{so}})^{-1} \gg a_{\text{ho}}$.

For two identical fermions, the $|T_0 \rangle$ channel experiences,
in general, the
same interaction as the $|S_0 \rangle$ channel. 
The interaction in the $|T_0 \rangle$ channel does, however, 
not
lead to any appreciable scattering
for short-range interactions tuned away from free-space
$p$-wave
resonances, 
implying that the interaction in the $|T_0 \rangle$
channel can be set to zero for most parameter combinations
without noticeable changes.
Section~\ref{sec_singlet} considers scattering properties
at the threshold
($E=E_{\text{th}}$)
while Sec.~\ref{sec_singlet_triplet} 
considers the above threshold behavior ($E > E_{\text{th}}$).

\subsection{Scattering properties at the lowest scattering threshold}
\label{sec_singlet}
This section compares
our finite-range results for $\underline{K}_{\text{phys}}$ and its
eigenvalues ${K}_{\text{phys}}^{(j)}$
with the
physical K-matrix $\underline{K}_{\text{1d,phys}}$ 
and its eigenvalues ${K}_{\text{1d,phys}}^{(j)}$,
which are obtained, as discussed in Sec.~\ref{sec_1dcoupling_soc},
by using $g_{\text{1d}}^{\text{soc}}(k_z)$ as input. We emphasize that
the energy dependence
of $g_{\text{1d}}^{\text{soc}}$, via the energy dependence of the
$s$-wave scattering length and the function $C$ [see Eq.~(\ref{eq_g1dsoc})], 
needs to be accounted for when comparing the results.

\begin{figure}[h]
\begin{center}
\includegraphics[width=2.8in]{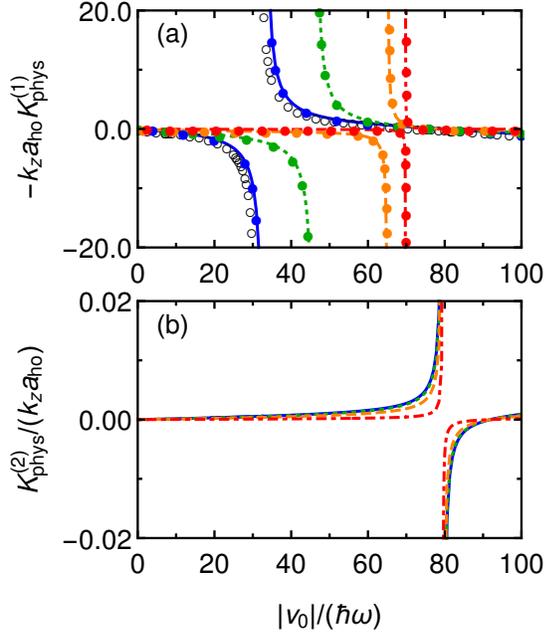}
  \caption{(color online)
Scattering properties for two identical fermions 
in the presence of spin-orbit coupling with  interaction in the singlet channel
only for $r_0=0.3 a_{\text{ho}} / \sqrt{2}$, 
$(k_{\text{so}})^{-1}=(0.2 \sqrt{2})^{-1} a_{\text{ho}}$,
and
$\tilde{\delta}=0$ (double-minimum regime).
Lines show the scaled eigenvalues
(a) $\lim_{k_z \rightarrow 0} [-k_z a_{\text{ho}} K_{\text{phys}}^{(1)}(k_z)]$
and 
(b) $\lim_{k_z \rightarrow 0} [K_{\text{phys}}^{(2)}(k_z) / (k_z a_{\text{ho}})]$,
obtained by diagonalizing the
    physical $2 \times 2$ K-matrix
    $\underline{K}_{\text{phys}}(k_z)$,
as a function of the magnitude 
of the depth $v_0$ of the Gaussian interaction potential. 
Blue solid, green dotted, orange dashed, and 
red dash-dotted lines are for 
$E_{\Omega}  = 0.01 \hbar \omega$,
$0.03 \hbar \omega$, $0.06 \hbar \omega$, and $0.079 \hbar \omega$
(corresponding to 
$E_{\Omega}=0.25 E_{\text{so}}, 0.75 E_{\text{so}}, 1.5 E_{\text{so}}$, 
and $1.975 E_{\text{so}}$), 
respectively.
In panel~(a), the open circles show the even-$z$ zero-range 
coupling constant $g_{\text{1d}}^{\text{even}}(k_z=0)$ 
[Eq.~(\ref{eq_g1deven_k1d}); spin-orbit coupling effects 
are not accounted for] and 
the filled circles show the
quantity
$\lim_{k_z \rightarrow 0} [-k_z a_{\text{ho}} K_{\text{1d,phys}}^{(1)}(k_z)]$.
    }
\label{fig4}
\end{center}
\end{figure}

We start our discussion by considering the double-minimum regime.
The lines in Fig.~\ref{fig4}
show the eigenvalues $K_{\text{phys}}^{(j)}$
of $\underline{K}_{\text{phys}}$ as a function of the
absolute value of the depth $v_0$ of the
two-body Gaussian potential with range $r_0=0.3 a_{\text{ho}}/ \sqrt{2}$ 
for various $\Omega$.
Motivated by the threshold law for the zero-range interactions
(see Table~\ref{tab_threshold}), Fig.~\ref{fig4}(a) shows the
quantity $\lim_{k_z \rightarrow 0} [-k_z a_{\text{ho}} K_{\text{phys}}^{(1)}(k_z)]$,
which can be interpreted as an {\em{effective}}
(dimensionless) even-parity coupling constant.
For $E_{\Omega} \ll E_{\text{so}}$,
the four spin channels 
are approximately decoupled and
the $k_{\text{so}}$
term can, in a lowest-order treatment, be gauged away.
As a consequence, the system properties are, to leading order,
expected to be identical to those obtained for the wave guide
in the absence of spin-orbit coupling
but with $s$-wave interactions.
The solid line in Fig.~\ref{fig4}(a) shows the scaled eigen value
for $E_{\Omega}=\hbar \omega/100=E_{\text{so}}/4$.
For this small $E_{\Omega}$,
the scaled eigen value $-k_z a_{\text{ho}} K_{\text{phys}}^{(1)}(k_z)$
is approximately
equal to $g_{\text{1d}}^{\text{even}}/(\hbar \omega a_{\text{ho}})$
[(dimensionless) even-parity coupling constant
in the absence of spin-orbit
coupling,
Eq.~(\ref{eq_g1deven_k1d}) with $a_s(E_{\text{th}})$; 
see open circles in Fig.~\ref{fig4}(a)].
This confirms that $-(\hbar^2 k_z / \mu) K_{\text{phys}}^{(1)}(k_z)$
behaves, at least in this small-$E_{\Omega}$ limit, like
an effective one-dimensional even-parity coupling constant
and that the system in the $E_{\Omega} \rightarrow 0$ limit
deviates only weakly from the system without spin-orbit coupling.
Figure~\ref{fig4}(a)
shows that the divergence of $-k_z a_{\text{ho}} K_{\text{phys}}^{(1)}(k_z)$
moves to
larger $|v_0|$ [corresponding to smaller $a_s(E_{\text{th}})$]
as $E_{\Omega}$ increases from $0.25 E_{\text{so}}$ (solid line) to $0.75 E_{\text{so}}$
(dashed line) to
$1.5 E_{\text{so}}$ (dotted line)
to
$1.975 E_{\text{so}}$ (dash-dotted line).
At the same time, the resonance becomes---this can be seen when the
data in Fig.~\ref{fig4}(a) are replotted as a function of 
the scattering length---narrower.
We attribute
the narrowing of the resonances with increasing $E_{\Omega}$
to the fact that the contribution of the $|S_0 \rangle$
channel to the lowest threshold
decreases with increasing $E_{\Omega}$.
The results for the finite-range potential
[lines in Fig.~\ref{fig4}(a)] are well reproduced by the results for the zero-range 
model that accounts for the spin-orbit
coupling [filled circles in Fig.~\ref{fig4}(a)].

\begin{figure}[h]
\begin{center}
\includegraphics[width=2.5in]{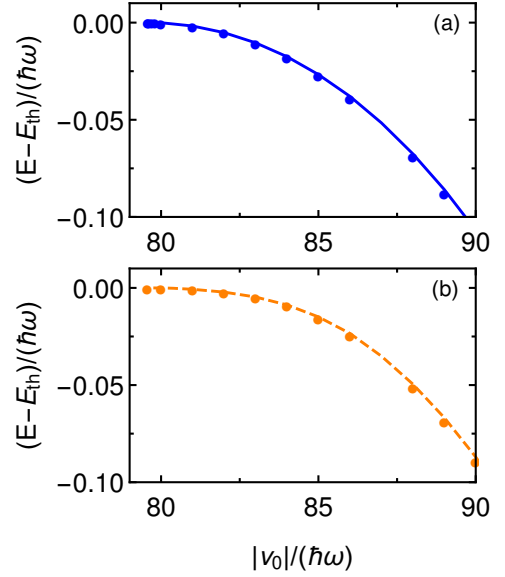}
  \caption{(color online)
Bound state properties for two identical fermions 
in the presence of spin-orbit coupling with interaction in the singlet channel
only for $r_0=0.3 a_{\text{ho}} / \sqrt{2}$, 
$(k_{\text{so}})^{-1}=(0.2 \sqrt{2})^{-1} a_{\text{ho}}$,
and
$\tilde{\delta}=0$ (double-minimum regime).
The solid lines show the energy of the bound state,
relative to the lowest scattering threshold,
as a function of the magnitude of the depth $v_0$
for
(a) $E_{\Omega}= \hbar \omega/100 = E_{\text{so}}/4$
and
(b) $E_{\Omega}= 3 \hbar \omega/50 = 3 E_{\text{so}}/2$.
For comparison, the filled circles show the zero-range 
energies,
relative to the lowest scattering threshold,
obtained by solving the eigen equation in the presence of
spin-orbit coupling self-consistently [see
Eq.~(27) of Ref.~\cite{PRA_Zhang}].
    }
\label{fig_bound_dm}
\end{center}
\end{figure}

The zero-range model predicts that the second eigenvalue of $\underline{K}_{\text{1d,phys}}(k_z)$
vanishes identically in the double-minimum regime for all energies
(see Table~\ref{tab_threshold}).
In constrast,
$K_{\text{phys}}^{(2)}(k_z)$
does not vanish for the finite-range interaction potential.
Specifically, we find that  
$K_{\text{phys}}^{(2)}(k_z)/(k_z a_{\text{ho}})$
approaches a constant in the small $k_z$ limit
[see lines in Fig.~\ref{fig4}(b)].
This scaling 
suggests that the quantity
$\lim_{k_z \rightarrow 0}[\hbar^2/(\mu k_z) K_{\text{phys}}^{(2)}(k_z)]$
can be interpreted
as an {\em{effective}} odd-$z$ coupling constant.
To elucidate this
interpretation, we 
focus on the well depth $v_0$,
at which $K_{\text{phys}}^{(2)}(k_z)/(k_z a_{\text{ho}})$
diverges ($|v_0| \approx 80 \hbar \omega$). Comparison with Fig.~\ref{fig_nosoc}(c)
shows that $K_{\text{phys}}^{(2)}(k_z)/(k_z a_{\text{ho}})$ diverges at approximately the same
$v_0$ at which the 
system without spin-orbit coupling supports an even-$z$ bound state
with infinitesimally small binding energy.
To calculate the bound state energies in the presence of the spin-orbit
coupling terms and the external wave guide confinement, we 
employ a basis set expansion 
approach~\cite{GB3b,Varga}.
Our results
(see lines in Fig.~\ref{fig_bound_dm})
show that the critical $v_0$,
at which a new two-body bound state
is first supported for the
$k_{\text{so}}$ considered ($\Omega < \Omega_*$), 
does not differ significantly 
from the case without spin-orbit coupling.
We
find that the bound state in the presence
of spin-orbit coupling contains, in general, 
both even- and odd-$z$ contributions.
For $\Omega < \Omega_*$, the triplet contribution is largest
when the binding energy is smallest. Correspondingly, 
the effective
odd-$z$ coupling constant  $[\hbar^2/(\mu k_z)] K_{\text{phys}}^{(2)}(k_z)$
diverges when a new bound state
is being pulled in.
For comparison,
the filled circles in Fig.~\ref{fig_bound_dm} show the
energy of the bound state predicted by the
zero-range calculations in the presence of
spin-orbit coupling [Eq.~(27) of Ref.~\cite{PRA_Zhang}]. 
The agreement with our finite-range calculations is good,
confirming the validity of both the finite- and zero-range bound state
calculations.
We find that the width of the resonance feature in  Fig.~\ref{fig4}(b)
decreases with decreasing $r_0$, suggesting that
the finiteness of $K_{\text{phys}}^{(2)}(k_z) / (k_z a_{\text{ho}})$ 
is, indeed, due to the finite range of the Gaussian interaction potential.

The physical K-matrix contains, 
in general, off-diagonal matrix elements, which reflect the fact
that the different spin channels are coupled and correspondingly that
there exists a coupling
between the even-$l$ and odd-$l$ partial wave channels
due to the presence of the spin-orbit coupling terms.
Diagonalizing the physical K-matrix, as done
to obtain the results 
shown in Fig.~\ref{fig4}, corresponds to changing the asymptotic basis
that the inner solution is
being matched to. While such a basis transformation 
provides---as illustrated above---useful insights, we emphasize that the 
full physical K-matrix
$\underline{K}_{\text{phys}}(k_z)$
is needed to determine, e.g., partial transmission and reflection
coefficients.

For $E_{\Omega} \gg E_{\text{so}}$ (single-minimum regime),
one can apply a rotation to the relative Hamiltonian such that it
is approximately block-diagonal.
Since the 
lowest scattering threshold for $\Omega > \Omega_*$
has only triplet contributions
and since the interactions in the triplet channels are set to zero
in this section,
one naively expects
that
scattering resonances would be absent. 
However, since there exists a coupling
between the different channels, 
the existence of 
a bound state due to the interaction in the $|S_0 \rangle$
channel
leads, as discussed next, to a scattering 
resonance,
provided the scattering energy is degenerate with the bound state
energy (see also Ref.~\cite{scientific_report}). 

\begin{figure}[h]
\begin{center}
\includegraphics[width=3.2in]{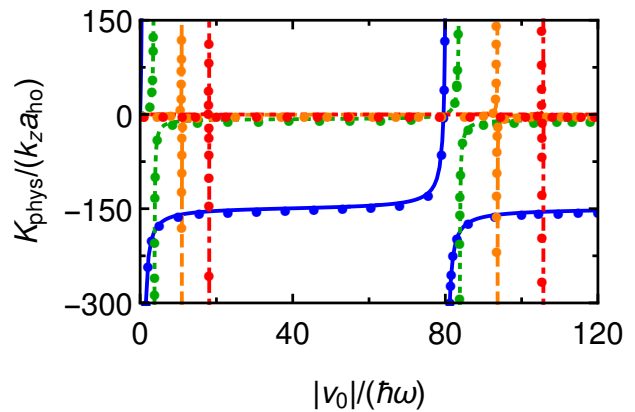}
  \caption{(color online)
Scattering properties for two identical fermions 
in the presence of spin-orbit coupling with interaction in the singlet channel
only for $r_0=0.3 a_{\text{ho}} / \sqrt{2}$, 
$(k_{\text{so}})^{-1}=(0.2 \sqrt{2})^{-1} a_{\text{ho}}$,
and $\tilde{\delta}=0$ (single-minimum regime).
Blue solid, green dotted, orange dashed, and 
red dash-dotted lines
show the scaled physical K-matrix
$\lim_{k_z \rightarrow 0} [{K}_{\text{phys}}(k_z)/ (k_z a_{\text{ho}})]$
as a function of the magnitude of the
depth $v_0$ of the Gaussian potential
for $E_{\Omega}  = 2.025 E_{\text{so}}$,
$2.5 E_{\text{so}}$, $7.5 E_{\text{so}}$, and $25 E_{\text{so}}$, respectively.
The filled circles show the quantity
$\lim_{k_z \rightarrow 0} [{K}_{\text{1d,phys}}(k_z)/ (k_z a_{\text{ho}})]$.
  }
\label{fig5}
\end{center}
\end{figure}

\begin{figure}[h]
\begin{center}
\includegraphics[width=2.8in]{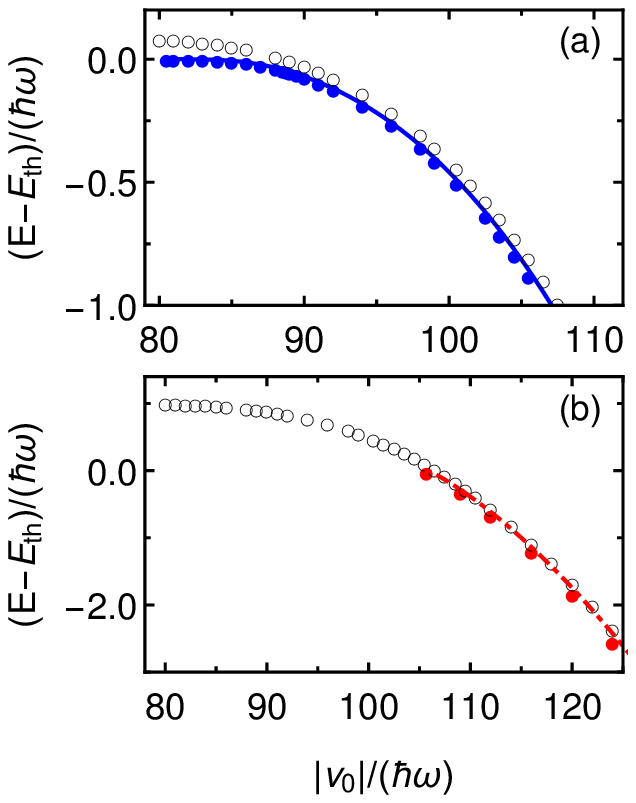}
  \caption{(color online)
Bound state energies for two identical fermions 
in the presence of spin-orbit coupling with interaction in the singlet channel
only for $r_0=0.3 a_{\text{ho}} / \sqrt{2}$, 
$(k_{\text{so}})^{-1}=(0.2 \sqrt{2})^{-1} a_{\text{ho}}$,
and
$\tilde{\delta}=0$
(single-minimum regime).
Solid and dash-dotted lines
show the energy of the bound state, relative to the lowest 
scattering threshold,
as a function of the magnitude of the 
depth $v_0$ of the Gaussian potential
for (a) $E_{\Omega}  = 2.025 E_{\text{so}}$
and (b) $E_{\Omega}  = 25 E_{\text{so}}$, respectively.
The filled circles and open circles show the
energy of the zero-range model in the presence and absence of
spin-orbit coupling terms; the 
zero-range eigen energy equations are solved self-consistently,
taking the energy-dependence of the $s$-wave scattering length
into account.
    }
\label{fig5a}
\end{center}
\end{figure}

Motivated by the threshold laws reported in
Table~\ref{tab_threshold}, the dash-dotted line in Fig.~\ref{fig5}
shows the scaled physical K-matrix 
$\lim_{k_z \rightarrow 0} [K_{\text{phys}}(k_z)/(k_z a_{\text{ho}})]$
for $E_{\Omega}= 25 E_{\text{so}}=\hbar \omega$ as a function of the 
magnitude of the depth $v_0$
of the two-body potential with $r_0=0.3 a_{\text{ho}} / \sqrt{2}$.
The results are in good 
agreement with the zero-range model predictions
(filled circles in Fig.~\ref{fig5}).
The dash-dotted line in Fig.~\ref{fig5}
diverges at $|v_0|=105.7 \hbar \omega$.
Figure~\ref{fig5a}(b) shows that this
  is the same $v_0$ at which 
the bound state becomes unbound (the dash-dotted
line shows the energy of the bound state, relative to
the energy of the lowest threshold, for the same parameters).
The energy of the bound state is reproduced reasonably well
by solving the eigen equation that does not
account for the spin-orbit coupling terms
self-consistently
for the even-$z$ zero-range
energy, using the energy-dependent $s$-wave
scattering length [open circles in Fig.~\ref{fig5a}(b)].
Even though the bound state becomes,
strictly speaking, unbound when its energy is above $E_{\text{th}}$,
we can think of the bound state as turning into a resonance 
or quasi-bound state for $E > E_{\text{th}}$.
The resonance state, in turn, becomes unbound when it hits the minimum
of the second-lowest dispersion curve (recall
that one of the two degenerate states associated
with this second-lowest scattering threshold is a pure $|S_0 \rangle$
state),
which sits $E_{\Omega}$ above $E_{\text{th}}$
(here, $E_{\Omega}=\hbar \omega$).
While we 
did not calculate the energy of this resonance state for the finite-range
Gaussian potential, we believe that the open circles shown in Fig.~\ref{fig5a}(b) 
provide a reasonable description of the energy of the resonance state.
As discussed further in Sec.~\ref{sec_singlet_triplet}
(see also Ref.~\cite{scientific_report}),
the resonance state leads to an above-threshold scattering resonance
(reflection coefficient of one in Fig.~\ref{fig_energy4}).
For comparison, the filled circles in Fig.~\ref{fig5a}(b)
show the energy of the bound state
predicted by Eq.~(27) of Ref.~\cite{PRA_Zhang} for the zero-range
interaction model in the presence of
spin-orbit coupling
(again, the equation is solved self-consistently).
The agreement with the dash-dotted line is reasonably good.

Figure~\ref{fig5} shows that the
resonance position shifts with decreasing $\Omega$.
Importantly, the resonance position coincides with the
position where the energy of the bound state in the presence of the
spin-orbit coupling terms hits the lowest
scattering threshold
for all $\Omega > \Omega_*$.
Since the deviations from the block-diagonal structure
of the rotated Hamiltonian 
increase with decreasing $\Omega$, the resonance becomes broader
(the broadness of the resonance is measured in terms of the width
in the scattering length; the corresponding plot is not shown).
This interpretation is consistent with our analysis of the
bound state wave function for an energy just below
the scattering threshold.
For $E_{\Omega}=25 E_{\text{so}}$ and $E_{\Omega}=2.025 E_{\text{so}}$, e.g.,
the bound state with binding energy $1.14 \times 10^{-4} \hbar \omega$
has a probability of about $21$~\% and $5$~\%, respectively, to be
in the $|S_0 \rangle$ channel.
Moreover, since the coupling between the different
spin channels increases with decreasing $\Omega$, the
bound state energy is, for smaller $\Omega$,
not overly well reproduced by the bound state
expression that does not account for the spin-orbit
coupling [open circles in Fig.~\ref{fig5a}(a)].
As expected,
the bound state energy is
well reproduced by the 
zero-range
expression from Ref.~\cite{PRA_Zhang} 
[filled circles in Fig.~\ref{fig5a}(a)],
provided the energy-dependence
of the scattering length is accounted for.

Figure~\ref{fig5} shows another interesting aspect.
Close to the transition from the single-minimum to the
double-minimum regime (the solid line is for 
$E_{\Omega} = 2.025 E_{\text{so}}$), the 
quantity $K_{\text{phys}}(k_z) / (k_z a_{\text{ho}})$
is negative and approximately constant for all $v_0$
away from the resonance.
This behavior is in agreement with the zero-range results.
Taylor expanding the coefficient
$a_{II,1}^{(+)}$ 
[see Eqs.~(\ref{eq_coeff_threshold}),
(\ref{eq_aroman21_plus}), and (\ref{eq_broman21_plus})],
which governs the near-threshold
behavior of $K_{\text{1d,phys}}$ in the single-minimum regime,
around $\Omega=\Omega_*$,
we find
\begin{align}
(a_{\text{ho}})^{-1} a_{II,1}^{(+)}=
-\frac{E_{\text{so}}}{E_\Omega-2E_{\text{so}}}\frac{1}{k_{\text{so}} a_{\text{ho}} }.
\end{align}
Correspondingly,
the quantity
$\lim_{k_z \rightarrow 0}K_{\text{1d,phys}}(k_z) / (k_z a_{\text{ho}})$
approaches negative infinity for all
interaction strengths when $\Omega$ approaches
$\Omega_*$ from above.
The fact that the scaled K-matrix goes
to minus infinity regardless of the details
of the underlying two-body potential
indicates that the 
physics is governed by the relative dispersion curves.
As $\Omega$ approaches $\Omega_*$, the bottom of the relative dispersion curve
becomes ``flatter'', thereby leading to a larger density of states or
degeneracy. In terms of the Bose-Fermi 
duality~\cite{BoseFermi1,GOBoseFermi}, 
one can interpret the
fact that the effective odd-$z$ coupling constant goes to infinity
as a signature of bosonization, which is facilitated by the
enhanced degeneracy~\cite{flatband}.

\begin{figure}[h]
\begin{center}
\includegraphics[width=2.8in]{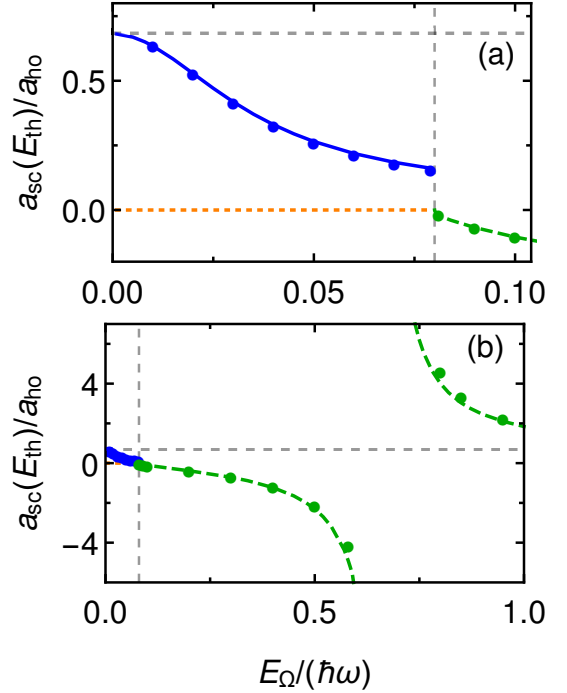}
  \caption{(color online)
Summary of resonance positions for two identical fermions
with interaction in the singlet channel only.
    The blue solid, orange dotted, and green dashed lines
    show the resonance positions, obtained for the Gaussian interaction
model, of
    $\lim_{k_z \rightarrow 0} [k_z a_{\text{ho}} \underline{K}_{\text{phys}}(k_z)]$ 
in the double-minimum regime,
    $\lim_{k_z \rightarrow 0} [k_z a_{\text{ho}} K_{\text{phys}}^{(2)}(k_z)]$ 
in the double-minimum regime,
    and
    $\lim_{k_z \rightarrow 0} [K_{\text{phys}}(k_z)/(k_z a_{\text{ho}})]$ 
in the single-minimum regime,
    respectively.
    For comparison, the filled circles show the zero-range
results, which are obtained by taking the energy-dependence
of $g_{\text{1d}}^{\text{soc}}$ into account.
Panels (a) and (b) cover different $E_{\Omega}$ ranges.
The grey dashed horizontal line marks the resonance 
position for the case where spin-orbit coupling 
is absent (even-$z$ solution).
The grey
dashed vertical line marks the transition from the double-minimum
to the single-minimum regime.
  }
\label{fig_res_position}
\end{center}
\end{figure}

Figure~\ref{fig_res_position} shows
the energy-dependent $s$-wave scattering length
$a_s(E_{\text{th}})$ at which the divergence
for the scattering energy $E=E_{\text{th}}$
and $(k_{\text{so}})^{-1} = (0.2 \sqrt{2})^{-1} a_{\text{ho}}$
occurs  as a function of $\Omega$.
The resonance positions are calculated for the Gaussian potential
with a relatively large
range $r_0$, namely $r_0=0.3 a_{\text{ho}} / \sqrt{2}$.
Repeating the calculations in selected cases for smaller $r_0$,
we find that the resonance occurs at roughly the same
$s$-wave scattering length, provided the
energy-dependent scattering lengths 
$a_s(E_{\text{th}})$ for two different
potentials are compared.
The solid, dotted, and dashed lines
show the resonance 
positions of $k_z a_{\text{ho}} \underline{K}_{\text{phys}}$ 
($\Omega < \Omega_*$),
${K}^{(2)}_{\text{phys}} / (k_z a_{\text{ho}})$ 
($\Omega < \Omega_*$),
and
${K}_{\text{phys}}  / (k_z a_{\text{ho}})$ 
($\Omega > \Omega_*$),
respectively.
The zero-range predictions (filled circles) agree well with the
results for the finite-range potential.
It can be seen that the resonance position can be tuned significantly 
by the spin-orbit coupling terms.

\subsection{Scattering properties as a function of the energy}
\label{sec_singlet_triplet}
This section extends the calculations presented in the previous section 
to scattering energies above the scattering threshold.
As in the previous section, we keep the transverse confinement
length $a_{\text{ho}}$ and the spin-orbit coupling strength
$k_{\text{so}}$ fixed and vary the Raman coupling strength $\Omega$
and the depth $v_0$ of the two-body potential [or, equivalently, the 
zero-energy $s$-wave scattering length $a_s(0)$].
Since the dimension of $\underline{K}_{\text{phys}}$ depends on the scattering
energy (the number of energetically open channels
increases with increasing energy, 
respectively, monotonically and non-monotonically
in the single-minimum and double-minimum regimes),
it is not convenient to
present the individual K-matrix elements. Instead, we present the total
reflection coefficient ${\cal{R}}$, which 
is obtained by combining the 
individual K-matrix elements 
[see Eqs.~(\ref{eq_scattering_amplitude}), (\ref{eq_calr}),
(\ref{eq_calr_partial}),
and (\ref{eq_calr_total})].
From a physical point of view, ${\cal{R}}$ tells one the fraction
of the incoming flux that is reflected, provided the
incoming flux populates the energetically open channels
equally. In selected cases, we discuss the decomposition of the total
reflection coefficient into the coefficients ${\cal{R}}_t$, which are calculated 
assuming that the incoming flux populates only the $t$-th channel.

The dash-dotted lines and filled circles
in Figs.~\ref{fig_energy1}(a)-\ref{fig_energy1}(d) 
show the total reflection coefficient 
${\cal{R}}$ as
a function of the scattering energy $E$ for 
$a_s(E=0) \approx 0.612 a_{\text{ho}}$ 
(corresponding to 
$r_0=0.3a_{\text{ho}}/\sqrt{2}$ and $|v_0|=30\hbar\omega$) and 
$\Omega=0.06\hbar\omega=1.5E_{\text{so}}$ (double-minimum regime).
The agreement between the dash-dotted lines, which are obtained 
using $\underline{K}_{\text{phys}}$
for the finite-range Gaussian
potential,
and the green filled circles, which are
obtained using the effective one-dimensional
low-energy Hamiltonian $H_{\text{1d}}$ (taking the
energy-dependence of the scattering length for the Gaussian potential into account),
is very good.
This indicates that the effective 
one-dimensional low-energy Hamiltonian 
provides a good description in this parameter regime, 
provided the energy-dependence
of the $s$-wave scattering length is accounted for.
The scattering energies at which the number of energetically open channels
changes 
are shown by thin vertical lines.
It can be seen that the total reflection coefficient or its derivative
change discontinuously at these scattering energies.
The behavior of the total reflection coefficient
${\cal{R}}$ near these scattering energies can be 
obtained in analytical form using the threshold behavior of the K-matrix elements
listed in Table~\ref{tab_threshold}.
For example, using the results from Table~\ref{tab_threshold}
in Eq.~(\ref{eq_connect_k_and_r}), one can analytically describe the behavior
of $\cal{R}$ near $E=E_{III}=\hbar \omega$ for $a_s \ne 0$.
We find that $\cal{R}$ scales as $(k_{III}^{(-)})^2$
as the scattering energy approaches $\hbar \omega$ from below
[this explains why $\cal{R}$ goes to zero as $E$ approaches $(E_{III})^-$ in
Fig.~\ref{fig_energy1}(b)]
and that $\cal{R}$ goes to one [left edge of Fig.~\ref{fig_energy1}(c)]
as the scattering energy approaches $(E_{III})^+$.
The jump of $\cal{R}$ from zero to one
at $E=E_{III}=\hbar \omega$ can be 
attributed to the opening of new channels as the energy changes from
below $\hbar \omega$ to above $\hbar \omega$.
The behavior of the total reflection coefficient 
$\cal{R}$ just below $E=2.9375 \hbar \omega$ is a bit different.
As $E$ approaches $2.9375 \hbar \omega$ from below, the reflection
coefficient does not go to zero but takes a 
value that depends on the system parameters. 
For the case at hand, the close-to-zero value 
of ${\cal{R}}$ just below $E=2.9375 \hbar \omega$
can be traced back to the suppression of ${\cal{R}}$ by
a small $g_{\text{1d}}^{\text{soc}}$. 

The dashed, dotted, and solid lines in Fig.~\ref{fig_energy1}
show the coefficients 
${\cal{R}}_1$, ${\cal{R}}_3$, and ${\cal{R}}_4$.
The sum of these coefficients, including only the energetically open channels,
yields ${\cal{R}}$.
It can be seen that the total reflection coefficient ${\cal{R}}$
contains appreciable contributions from multiple coefficients ${\cal{R}}_t$
in the cases where more than one channel is energetically open
[Figs.~\ref{fig_energy1}(a), \ref{fig_energy1}(c), and \ref{fig_energy1}(d)].
The reason that multiple ${\cal{R}}_t$ contribute
is a consequence of the fact that the spin-orbit coupling
terms in the double-minimum
regime induce,
in general, a non-perturbative coupling between the singlet and triplet
spin states.
As discussed in Sec.~\ref{sec_physical}, the sum of ${\cal{R}}$ and ${\cal{T}}$ is
equal to the number $N^{\text{o}}$
of energetically open channels, i.e., equal to 
2, 1, 3, and 4 for Figs.~\ref{fig_energy1}(a), \ref{fig_energy1}(b),
\ref{fig_energy1}(c), and \ref{fig_energy1}(d), respectively.
Thus, the system is ``fully transparent'' for
scattering energies $E \lesssim \hbar \omega$,
``nearly fully transparent'' for
$E \lesssim 2.9375\hbar \omega$ (see above), 
and 
``fully reflective'' for a scattering energy 
a bit larger than $0.94 \hbar \omega$
(but not for $E=E_{\text{th}}$).

\begin{widetext}

\begin{figure}[h!]
\begin{center}
\includegraphics[width=6.6in]{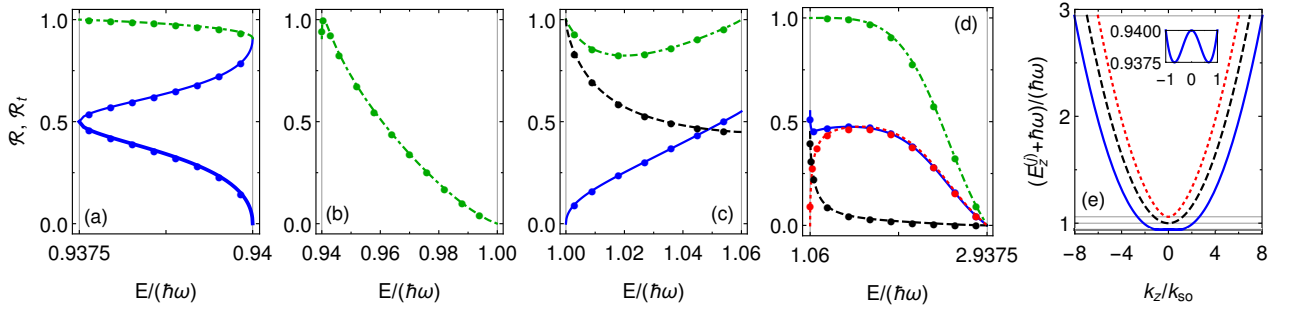}
  \caption{(color online)
Scattering observables for two identical fermions
with interaction in the singlet channel
only as a function of the scattering energy
$E$ for $(k_{\text{so}})^{-1}=(0.2\sqrt{2})^{-1}a_{\text{ho}}$ 
and $E_\Omega =0.06\hbar\omega=1.5E_{\text{so}}$ (double-minimum regime).
The number of energetically open channels is two in panel~(a),
one in panel~(b),
three in panel~(c), and four in panel~(d).
The green dash-dotted lines
show the total reflection coefficient 
${\cal{R}}$ obtained by applying the full K-matrix formalism to the
Gausssian potential with $r_0=0.3 a_{\text{ho}}/\sqrt{2}$ and 
$|v_0|=30 \hbar \omega$
[$a_s(0) \approx 0.612 a_\text{ho}$].
For comparison, the filled circles show ${\cal{R}}$ obtained 
by using the effective
one-dimensional low-energy Hamiltonian $H_{\text{1d}}$
(the effective coupling constant
$g_{\text{1d}}^{\text{soc}}$ 
is determined using the energy-dependence of $a_s$ for the Gaussian potential).
The black dashed, red dotted, thin blue solid, 
and thick blue solid lines show the quantities
${\cal{R}}_1$, ${\cal{R}}_3$, ${\cal{R}}_{4,+}$
(solution for positive $k_z$), and ${\cal{R}}_{4,-}$
(solution for negative $k_z$), respectively,
obtained using the full K-matrix formalism. 
${\cal{R}}_2$  (not shown) is identically zero.
The filled circles (same color coding) are obtained 
by using $H_{\text{1d}}$.
The energies at which the number of energetically
open scattering channels change
are shown by thin vertical lines.
Panel~(e) shows 
the corresponding non-interacting relative dispersion curves.
The energies at which
the number of energetically open channels changes
are marked by solid horizontal lines.
The inset shows an enlargement of the lowest dispersion
curve.
  }
\label{fig_energy1}
\end{center}
\end{figure}

To understand the dependence of the total reflection coefficient on the scattering
length, we analyze $\underline{K}_{\text{1d,phys}}$,
obtained from the effective low-energy Hamiltonian for the 
zero-range potential,
for the same $k_{\text{so}}$ and $\Omega$
as those considered in Fig.~\ref{fig_energy1}. 
Figures~\ref{fig_energy2}(a)-\ref{fig_energy2}(d)
show contour plots of the reflection coefficient ${\cal{R}}$ as functions
of 
the $s$-wave scattering length $a_s$
and
the scattering energy $E$ 
[the energy regions considered are the same as in 
Figs.~\ref{fig_energy1}(a)-\ref{fig_energy1}(d)].
The reflection coefficient shows an appreciable scattering length dependence.
Interestingly, 
the reflection coefficient ${\cal{R}}$ approaches one at the
lowest scattering threshold for all $s$-wave scattering lengths,
except for $a_s=0$.
This is analogous to the situation without
spin-orbit coupling (see Fig.~1 of Ref.~\cite{Olshanii1998}),
with the difference that the total transmission coefficient
${\cal{T}}$ is also equal to one in the spin-orbit coupling case
(recall, we are considering the regime where $N^{\text{o}}=2$)
while it is zero in the absense of spin-orbit coupling
(in this case, $N^{\text{o}}=1$).
If one prepared the system in such a way that initially
only the rotated state corresponding to the 
eigen value $K_{\text{phys,1d}}^{(1)}$
was occupied, one should observe ``true'' full reflection.
While this situation
might be challenging to realize experimentally, 
considerations like this one help to understand the implications of the results
shown in Figs.~\ref{fig_energy1} and \ref{fig_energy2}.
%It is also interesting that the two-fermion system becomes fully
%transparent (${\cal{R}}=0$) at the right edges of
%Figs.~\ref{fig_energy2}(b) and \ref{fig_energy2}(d) 
%for almost all 
%$s$-wave scattering lengths.
As already discussed above, the two-fermion system becomes fully
transparent (${\cal{R}}=0$) at the right edge of
Fig.~\ref{fig_energy2}(b) and nearly
fully transparent at the right edge of Fig.~\ref{fig_energy2}(d) 
for almost all 
$s$-wave scattering lengths.

\begin{figure}[h]
\begin{center}
\includegraphics[width=6.6in]{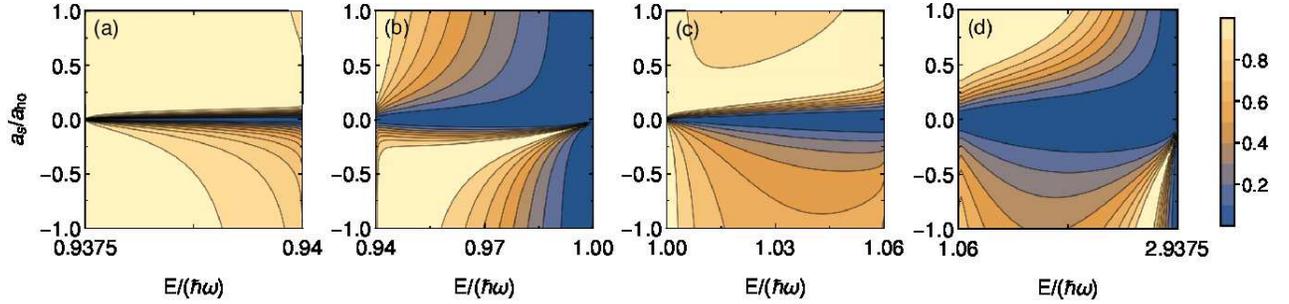}
  \caption{(color online)
Contour plot showing the
total reflection coefficient ${\cal{R}}$ for two identical fermions
with interactions in the singlet channel
only as functions of the $s$-wave scattering length $a_s$
and the scattering energy
$E$ 
for $(k_{\text{so}})^{-1}=(0.2\sqrt{2})^{-1}a_{\text{ho}}$
and $E_\Omega =0.06\hbar\omega=1.5E_\text{so}$ (double-minimum regime).
The results are obtained using the one-dimensional
 effective low-energy Hamiltonian $H_{\text{1d}}$
($a_s$ is treated as an input parameter).
  }
\label{fig_energy2}
\end{center}
\end{figure}

We now turn to the single-minimum regime.
Figures~\ref{fig_energy3} and \ref{fig_energy4} mirror
Figs.~\ref{fig_energy1} and \ref{fig_energy2}
using $E_\Omega=1.2\hbar\omega=30E_\text{so}$ instead of $E_\Omega=0.06\hbar\omega=1.5E_\text{so}$.
Because of the large $\Omega$, the highest
relative non-interacting dispersion curve with $n_{\rho}=0$
lies above the lowest 
relative non-interacting dispersion curve with $n_{\rho}=1$
[see Fig.~\ref{fig_energy3}(c)].
Since the coupling constant $g_{\text{1d}}^{\text{soc}}$,
Eq.~(\ref{eq_g1dsoc}),
is derived assuming that all $n_{\rho} \ge 1$ channels are closed,
the effective one-dimensional low-energy Hamiltonian $H_{\text{1d}}$
is only
valid for $E_{\text{th}} \le E \le 1.8\hbar\omega$ even though the 
scattering threshold of the highest $n_{\rho}=0$ channel lies
at an energy of $2.2\hbar\omega$.
Correspondingly, results for the effective low-energy
Hamiltonian are only shown in Fig.~\ref{fig_energy3}(a)
and not in Fig.~\ref{fig_energy3}(b).
As expected,
the dash-dotted line and the filled circles in Fig.~\ref{fig_energy3}(a)
agree well. Unlike in the double-minimum regime discussed in
Fig.~\ref{fig_energy1}, the total reflection coefficient
${\cal{R}}$ shown in Fig.~\ref{fig_energy3}
is dominated by a single channel, namely by ${\cal{R}}_1$.

Interestingly,
in the regime where only one channel is energetically
open [Fig.~\ref{fig_energy3}(a), $E \le \hbar \omega$], the system is nearly fully transparent.
This can be intuitively understood by realizing
that the singlet channel
contribution to the lowest scattering threshold vanishes (see Fig.~\ref{fig_threshold}).
Figure~\ref{fig_energy4}, which shows the total reflection
coefficient ${\cal{R}}$---obtained using $H_{\text{1d}}$---as 
functions of $a_s$ and $E$, confirms this. The
total reflection coefficient is zero or close to zero
for nearly
all scattering lengths, provided the scattering energy lies
between $E_{II}=-0.2\hbar \omega$ and $E_{III}= \hbar \omega$
[as already discussed above in the context of the
double-minimum case, $\cal{R}$
approaches one as $E \rightarrow (E_{III})^{-}$,
provided $a_s$ is not equal to $0$].
The total reflection coefficient 
in Fig.~\ref{fig_energy3} 
approaches one for specific $a_s$.
At these $a_s$,
the system supports a resonance state
(see Sec.~\ref{sec_singlet}).
When the scattering energy is equal to the energy of the
resonance state, the incoming flux gets 
reflected~\cite{scientific_report}.

\begin{figure}[h]
\begin{center}
\includegraphics[width=6.6in]{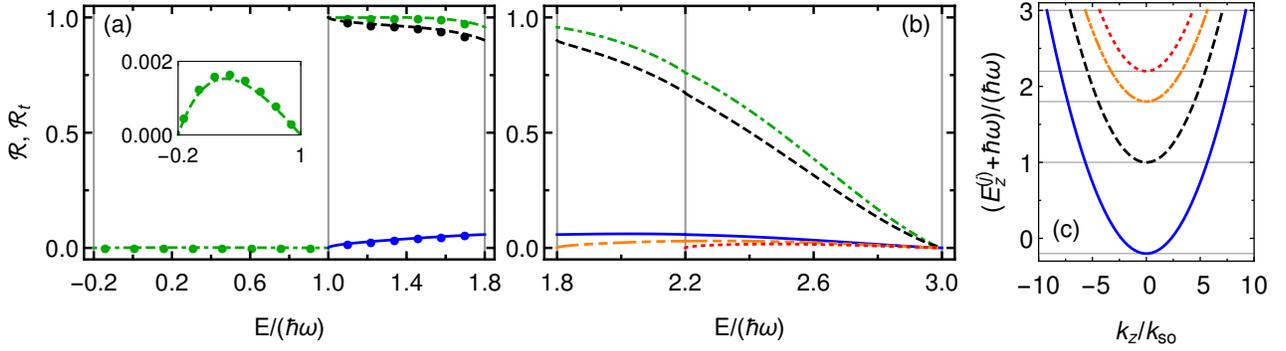}
  \caption{(color online)
Scattering observables for two identical fermions
with interactions in the singlet channel
only as a function of the scattering energy
$E$ for $(k_{\text{so}})^{-1}=(0.2\sqrt{2})^{-1}a_{\text{ho}}$
and $\Omega =1.2\hbar\omega=30E_\text{so}$ (single-minimum regime).
The green dot-dashed lines show the total reflection coefficient 
${\cal{R}}$ obtained by applying the full K-matrix formalism to the
Gaussian potential with $r_0=0.3 a_{\text{ho}}/\sqrt{2}$ and $|v_0|=30 \hbar \omega$
[$a_s(0) \approx 0.612a_{\text{ho}}$].
The black dashed, red dotted, and blue solid lines show the quantities
${\cal{R}}_1$, ${\cal{R}}_3$, and ${\cal{R}}_4$, respectively, 
obtained using the full K-matrix formalism. 
${\cal{R}}_2$ (not shown) is identically zero. 
The orange dash-dot-dotted line shows 
the contribution to the total reflection coefficient
that comes from flux
entering in the lowest relative dispersion curve with $n_\rho=1$.
For comparison, the filled circles in panel~(a) show the corresponding results
obtained using $\underline{K}_{\text{phys,1d}}$; the effective coupling constant
$g_{\text{1d}}^{\text{soc}}$ 
is obtained using the energy-dependence of $a_s$ for the Gaussian potential.
The energy regime covered in panel~(b)
is beyond the applicability regime of the effective
low-energy Hamiltonian.
The thin vertical lines mark the energies at which the
number of energetically open channels changes.
The inset in panel~(a) shows the same data as the 
main panel, but on an enlarged scale.
Panel~(c) shows 
the corresponding non-interacting relative dispersion curves.
The blue solid, black dashed, and red dotted
lines are for $n_{\rho}=0$ 
while the dash-dot-dotted line is for
$n_{\rho}=1$.
The energies at which 
the number of energetically open channels changes
are marked by horizontal solid lines.
  }
\label{fig_energy3}
\end{center}
\end{figure}  

\end{widetext}

\begin{figure}[h]
\begin{center}
\includegraphics[width=2.8in]{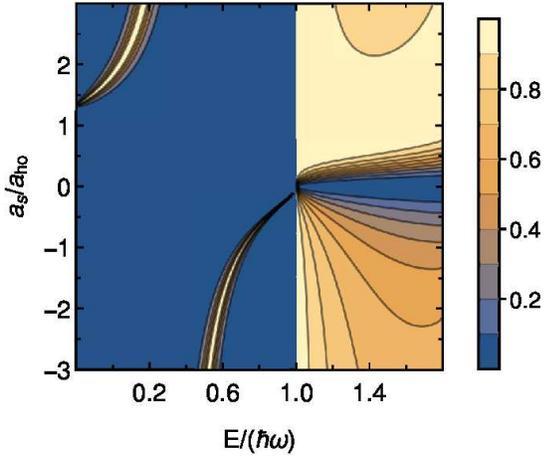}
  \caption{(color online)
Contour plot of total reflection coefficient ${\cal{R}}$ for two identical fermions
with interaction in the singlet channel
only as functions of the $s$-wave scattering length $a_s$
and the scattering energy
$E$ 
for $(k_{\text{so}})^{-1}=(0.2\sqrt{2})^{-1}a_\text{ho}$
and $\Omega =1.2\hbar\omega=30E_\text{so}$ (single-minimum regime).
The results are obtained using the one-dimensional
 effective low-energy Hamiltonian
$H_{\text{1d}}$ ($a_s$ is treated as an input parameter).
  }
\label{fig_energy4}
\end{center}
\end{figure}

\section{Conclusions}
\label{sec_conclusion}
This paper formulated the K-matrix scattering 
theory for two particles in effectively
one-dimensional space, realized by a tight wave guide
confinement,  in the presence of one-dimensional
spin-orbit coupling terms and applied it to two
identical fermions. The results for finite-range interactions were 
compared with results for zero-range interactions, which utilized 
an effective coupling constant that results from integrating out
the excited transverse wave guide modes, from the literature as input.
It was shown that the resonance positions can be
tuned by the spin-orbit coupling parameters. A parameter window was identified 
in which the scattering observables are essentially independent
of the underlying two-body potential.

The formulation and results presented provide the starting point for future studies.
It will be interesting to explore the case where the interactions in the
triplet channels contribute or even dominate.
It will also be interesting to apply the formalism to bosons.
One question concerns the construction of an effective low-energy single-band 
Hamiltonian based on the effective one-dimensional coupling constants
discussed in this work.
 Last, it will be interesting to extend the study to the three-body sector
and to two-body systems with different effective spin.

\section{Acknowledgement}
\label{acknowledgement}
We would like to thank Wei Zhang for discussions.
Support by the National Science Foundation through
grant numbers
PHY-1509892 and PHY-1745142
is gratefully acknowledged.
This work used the Extreme Science and Engineering
Discovery Environment (XSEDE), which is supported by
NSF Grant No. OCI-1053575,
and the OU
Supercomputing Center for Education and Research
(OSCER) at the University of Oklahoma (OU).

\appendix

\section{Generalized log-derivative algorithm}
\label{appendix_algorithm}
\subsection{Rewriting the coupled equations}
The Schr\"odinger equation
given in
Eq.~(\ref{eq_SEmatrix}) can be rewritten in the generic form
\begin{align}
\label{eq_SEmatrix_explicit}
\bigg[\underline{I}\frac{d^2}{dz^2}- 
\imath \underline{\alpha} \frac{d}{dz}+\underline{\beta}(z)\bigg]\underline{\phi}^{(m_l)}(z)=0,
\end{align}
where $\underline{I}$ denotes the $4 n_{\text{max}} \times 4 n_{\text{max}}$
identity matrix, 
\begin{align}
  \underline{\alpha}=\left(
  \begin{array}{cccc}
    \underline{a} & \underline{0} & \cdots & \underline{0} \\
    \underline{0} & \underline{a} &  &  \\
    \vdots &  & \ddots &  \\
    \underline{0} &  &  & \underline{a}
    \end{array}
  \right) ,
\end{align}
and
\begin{align}
  \underline{\beta}(z)=\left(
  \begin{array}{cccc}
    \underline{b}_{0,0}(z) & \underline{b}_{0,1}(z) & \cdots & \underline{b}_{0,n_{\text{max}}-1}(z) \\
    \underline{b}_{1,0}(z) & \underline{b}_{1,1}(z) &  &  \\
    \vdots &  & \ddots &  \\
    \underline{b}_{n_{\text{max}}-1,0}(z) &  &  & \underline{b}_{n_{\text{max}}-1,n_{\text{max}}-1}(z)
    \end{array}
  \right) .
\end{align}
The $4 \times 4$ matrices $\underline{a}$ and
$\underline{b}_{n_{\rho}',n_{\rho}}(z)$ are given by
\begin{align}
  \underline{a}= k_{\text{so}} 
\left(
  \begin{array}{cccc}
    0 & 0 & 0 & -2 \\
    0 & 0 & 0 & 0  \\
    0 & 0 & 0 & 0 \\
    -2 & 0 & 0 & 0
    \end{array}
  \right) 
\end{align}
and
\begin{widetext}
\begin{align}
  \underline{b}_{n_{\rho}',n_{\rho}}(z)= &
\frac{m}{\hbar^2} \left(
  \begin{array}{cccc}
    E-\epsilon_{n_{\rho},m_l} & 0 & 0 & 0 \\
    0 & E-\epsilon_{n_{\rho},m_l}-E_{\tilde{\delta}} & 0 & -E_{\Omega} / \sqrt{2}  \\
    0 & 0 & E-\epsilon_{n_{\rho},m_l}+E_{\tilde{\delta}} & -E_{\Omega}/ \sqrt{2} \\
    0 & -E_{\Omega} / \sqrt{2} & -E_{\Omega}/ \sqrt{2} & E-\epsilon_{n_{\rho},m_l}
    \end{array}
  \right) \delta_{n_{\rho}',n_{\rho}} - \nonumber \\
&  \frac{m}{\hbar^2} \left(
  \begin{array}{cccc}
    {\cal{V}}_{\text{int}}^{n_{\rho}',n_{\rho},S_0}(z) & 0 & 0 & 0 \\
    0 & {\cal{V}}_{\text{int}}^{n_{\rho}',n_{\rho},T_{+1}}(z) & 0 & 0  \\
    0 & 0 & {\cal{V}}_{\text{int}}^{n_{\rho}',n_{\rho},T_{-1}}(z) & 0 \\
    0 & 0 & 0 & {\cal{V}}_{\text{int}}^{n_{\rho}',n_{\rho},T_0}(z)
    \end{array}
  \right) .
\end{align}
\end{widetext}
Dividing $\underline{\phi}^{(m_l)}(z)$ into its
real and imaginary parts,
\begin{align}
\underline{\phi}^{(m_l)}(z) = \underline{\phi}^{(m_l)}_{\text{re}}(z) + 
\imath \underline{\phi}^{(m_l)}_{\text{im}}(z),
\end{align}
Eq.~(\ref{eq_SEmatrix_explicit})
can be rewritten as a purely real matrix equation:
\begin{align}
\label{eq_SEmatrix_explicit_double}
\bigg[\underline{I}\frac{d^2}{dz^2}+
\underline{A}(z) \frac{d}{dz}+\underline{B}(z)\bigg]
\underline{\varphi}^{(m_l)}(z)
=0.
\end{align}
Here, $\underline{I}$ is of size
$8 n_{\text{max}} \times 8 n_{\text{max}}$,
\begin{align}
\underline{A}(z)= \left(
  \begin{array}{cc}
    \underline{0} & \underline{\alpha} \\
    -\underline{\alpha} & \underline{0} 
    \end{array}
\right),
\end{align}
\begin{align}
\underline{B}(z)= \left(
  \begin{array}{cc}
    \underline{\beta} & \underline{0} \\
    \underline{0} & \underline{\beta} 
    \end{array}
\right),
\end{align}
and
\begin{align}
\label{eq_varphi_wave}
\underline{\varphi}^{(m_l)}(z)=
\left(
\begin{array}{cc}
\underline{\phi}^{(m_l)}_{\text{re}}(z) & \underline{\phi}^{(m_l)}_{\text{re}}(z) \\
\underline{\phi}^{(m_l)}_{\text{im}}(z) & \underline{\phi}^{(m_l)}_{\text{im}}(z) 
\end{array}
\right).
\end{align}
In writing Eq.~(\ref{eq_SEmatrix_explicit_double})
[see also Eq.~(\ref{eq_varphi_wave})],
we ``doubled'' the solution, i.e., the real part
$\underline{\phi}_{\text{re}}^{(m_l)}(z)$
and the imaginary part 
$\underline{\phi}_{\text{im}}^{(m_l)}(z)$
both appear twice.
In our case, $\underline{A}$ is independent of $z$.
We note, however, that the manipulations and algorithm outlined below
are also valid if $\underline{A}$ depends on $z$,
provided $\underline{A}^T$ is equal to $- \underline{A}$~\cite{SymmGLD}.
To emphasize this, we formally indicate the $z$-dependence of $\underline{A}$
in what follows.

Our goal is to propagate $\underline{\varphi}^{(m_l)}(z)$
from $z_{\text{min}}$ to $z_{\text{max}}$,
subject to appropriately chosen boundary conditions 
at $z_{\text{min}}$.
Since we are using Gaussian interaction potentials, the propagation starts
at $z_{\text{min}}=0$.
We write
\begin{align}
  \label{eq_bc1}
\underline{\phi}_{\text{re}}^{(m_l)}(0)=
\left(
  \begin{array}{cccc}
    \underline{\gamma}_{1} & \underline{0} & \cdots & \underline{0} \\
    \underline{0} & \underline{\gamma}_{1} &  &  \\
    \vdots &  & \ddots &  \\
    \underline{0} &  &  & \underline{\gamma}_{1}
    \end{array}
  \right), 
\end{align}
\begin{align}
\left( \frac{d}{dz}\underline{\phi}_{\text{re}}^{(m_l)}(z) \right) \Bigg| _
{z=0}=
\left(
  \begin{array}{cccc}
    \underline{\gamma}_{2} & \underline{0} & \cdots & \underline{0} \\
    \underline{0} & \underline{\gamma}_{2} &  &  \\
    \vdots &  & \ddots &  \\
    \underline{0} &  &  & \underline{\gamma}_{2}
    \end{array}
  \right) ,
\end{align}
\begin{align}
\underline{\phi}_{\text{im}}^{(m_l)}(0)=\underline{0},
\end{align}
and
\begin{align}
\left( \frac{d}{dz}\underline{\phi}_{\text{im}}^{(m_l)}(z) \right) \Bigg| _
{z=0} = \underline{0}.
\end{align}
The matrices $\underline{\gamma}_1$
and $\underline{\gamma}_2$ are chosen so that the total wave
function has the desired exchange symmetry.
For two identical fermions and even $m_l$
quantum number, e.g., the anti-symmetry of the
total wave function is fulfilled if we set
\begin{align}
  \underline{\gamma}_{1}= 
\left(
  \begin{array}{cccc}
    1 & 0 & 0 & 0 \\
    0 & 0 & 0 & 0  \\
    0 & 0 & 0 & 0 \\
    0 & 0 & 0 & 0
    \end{array}
  \right) 
\end{align}
and
\begin{align}
    \label{eq_bc6}
  \underline{\gamma}_{2}= 
\left(
  \begin{array}{cccc}
    0 & 0 & 0 & 0 \\
    0 & 1 & 0 & 0  \\
    0 & 0 & 1 & 0 \\
    0 & 0 & 0 & 1
    \end{array}
  \right) .
\end{align}
For a two-body potential with repulsive core, $\underline{\phi}^{(m_l)}_{\text{re}}$
and $d\underline{\phi}^{(m_l)}_{\text{re}}/dz$
would be set to $\underline{0}$ and $\underline{I}$, respectively,
at $z=z_{\text{min}}$.

\subsection{Formalism behind the algorithm}
It is useful to define the
propagators $\underline{L}^{(j)}(z',z'')$ with $j=1-4$,
\begin{align}
\label{eq_defLmatrix}
\begin{pmatrix}
    \underline{\varphi}'(z') \\
    \underline{\varphi}'(z'') 
\end{pmatrix}=
\begin{pmatrix}
    \underline{L}^{(1)} (z',z'') & \underline{L}^{(2)}  (z',z'')    \\
     \underline{L}^{(3)} (z',z'')  &  \underline{L}^{(4)}  (z',z'') 
\end{pmatrix}
\begin{pmatrix}
    \underline{\varphi} (z')      \\
     \underline{\varphi}(z'') 
\end{pmatrix},
\end{align}
where we introduced the abbreviations
$\underline{\varphi}(z')= {\underline{\varphi}}^{(m_l)}(z')$
and
\begin{align}
\underline{\varphi} '(z')=
\frac{d \underline{\varphi}^{(m_l)}(z)}{dz} \Big|_{z=z'}.
\end{align}
We rearrange Eq.~(\ref{eq_defLmatrix}) such that
the wave function and its derivative
at $z''$ can be, provided the propagators $\underline{L}^{(j)}(z',z'')$
are known, determined from the wave function
and its derivative at $z'$:
\begin{widetext}
\begin{align}
\label{eq_OmegarelationB2}
\begin{pmatrix}
    \underline{\varphi} (z'')      \\
     \underline{\varphi}'(z'') 
\end{pmatrix}
&=\begin{pmatrix}
-[\underline{L}^{(2)}(z',z'')]^{-1} \underline{L}^{(1)}(z',z'') & 
[\underline{L}^{(2)}(z',z'')]^{-1} \\
-\underline{L}^{(4)}(z',z'') [\underline{L}^{(2)}(z',z'')]^{-1} 
\underline{L}^{(1)}(z',z'')+\underline{L}^{(3)}(z',z'') &  
\underline{L}^{(4)}(z',z'') [{\underline{L}^{(2)}}(z',z'')]^{-1}
\end{pmatrix}
\begin{pmatrix}
    \underline{\varphi} (z')      \\
     \underline{\varphi}'(z') 
\end{pmatrix}.
\end{align}
\end{widetext}
The task is thus to find expressions for 
$\underline{L}^{(j)}(z',z'')$.

In the following we discuss the transformation used to
express the $\underline{L}^{(j)}(z',z'')$.
Following Ref.~\cite{SymmGLD}, we employ the transformation
\begin{align}
  \label{eq_changebasis}
\underline{\varphi}^{(m_l)}(z)
= 
{\underline{\cal{T}}}(z,\bar{z}) \underline{\varphi}^{(m_l)}_{\bar{z}}(z)
\end{align}
to remove the first derivative 
with respect to $z$ from Eq.~(\ref{eq_SEmatrix_explicit_double}).
Equation~(\ref{eq_changebasis})
can be interpreted as switching from an adiabatic basis to
a diabatic basis at each $z$.
Demanding that the identities
\begin{align}
\label{eq_TDE}
\bigg(\underline{I}\frac{d}{dz} + \frac{1}{2} \underline{A}(z)  \bigg)  
\underline{\mathcal{T}}(z,\bar{z})= \underline{0} 
\end{align}
and
\begin{align}
\label{eq_TDEa}
\underline{\mathcal{T}}(\bar{z},\bar{z})=\underline{I}
\end{align}
hold,
Eq.~(\ref{eq_SEmatrix_explicit_double}) becomes
\begin{align}
\label{eq_DEsim}
\nonumber
\bigg[ \frac{d^2 \underline{\mathcal{T}}(z,\bar{z}) }{dz^2}  
+  
\underline{\mathcal{T}}(z,\bar{z}) \frac{d^2}{dz^2}+ 
\underline{A}(z) 
\frac{d \underline{\mathcal{T}}(z,\bar{z})}{dz} +
\\
\underline{B}(z)  
\underline{\mathcal{T}}(z,\bar{z}) \bigg]
\underline{\varphi}^{(m_l)}_{\bar{z}}(z)=0.
\end{align}
Multiplying Eq.~(\ref{eq_DEsim}) from the left with
$\underline{\mathcal{T}}^T(z,\bar{z})$,
using Eq.~(\ref{eq_TDE}), and using that $\underline{A}(z)=-[\underline{A}(z)]^T$,
we obtain
\begin{align}
\left[ \underline{I} \frac{d^2}{d z^2} + \underline{B}_{\bar{z}}(z) \right]
\underline{\varphi}^{(m_l)}_{\bar{z}}(z)= 0,
\end{align}
where
\begin{align}
  \underline{B}_{\bar{z}}(z) =
\underline{\mathcal{T}}^T(z,\bar{z})
\underline{V}_{\text{eff}}(z)
\underline{\mathcal{T}}(z,\bar{z})
\end{align}
and
\begin{align}
\underline{V}_{\text{eff}}(z) =
\underline{B}(z) - \frac{1}{4} \underline{A}(z)  \underline{A}(z) 
-\frac{1}{2} \frac{d \underline{A}(z)}{dz}.
\end{align}

In our case, $d \underline{A}(z)/dz$ is equal to zero
and Eqs.~(\ref{eq_TDE}) and (\ref{eq_TDEa})
can be solved analytically:
\begin{align}
\underline{\mathcal{T}}(z,\bar{z})=
\frac{-\sin \left(k_{\text{so}}(z-\bar{z})\right) }{2 k_{\text{so}}} 
\underline{A}
+
\left(
  \begin{array}{cccc}
    \underline{t}_{D} & \underline{0} & \cdots & \underline{0} \\
    \underline{0} & \underline{t}_{D} &  &  \\
    \vdots &  & \ddots &  \\
    \underline{0} &  &  & \underline{t}_{D}
    \end{array}
  \right) 
,
\end{align}
where
\begin{align}
  \underline{t}_D =
\left(
  \begin{array}{cccc}
    \cos \left(k_{\text{so}}(z-\bar{z})\right) & 0 & 0 & 0 \\
    0 & 1 & 0 & 0  \\
    0 & 0 & 1 & 0 \\
    0 & 0 & 0 & \cos \left(k_{\text{so}}(z-\bar{z})\right)
    \end{array}
  \right) .  
  \end{align}
The algorithm discussed in the next section is based on the fact
that the transformation from 
$\underline{\varphi}^{(m_l)}(z)$
to $\underline{\varphi}^{(m_l)}_{\bar{z}}(z)$ 
can be performed at each $z$.

\subsection{Step by step algorithm}
To perform the propagation of
$\underline{\varphi}^{(m_l)}(z)$ and its derivative, we divide the interval
$[z_{\text{min}},z_{\text{max}}]$ into $N$ sectors
of length $2 h$. The grid points
are labeled $z_l$, where $l$ takes the values $0, 2,\cdots,{2N}$.
Note, however, 
that the algorithm also uses the ``midpoints'' $z_1,z_3,\cdots$.
The algorithm is formulated in terms of a number of 
auxiliary quantities:
\begin{align}
\underline{B}_{z_{l+2}}(z_{l+1})=\underline{\mathcal{T}}^T(z_{l+1},z_{l+2})
\underline{V}_{\text{eff}}(z_{l+1}) \underline{\mathcal{T}}(z_{l+1},z_{l+2}),
\end{align}
\begin{align}
\underline{s}_{l,l+2}= \bigg[   \frac{1}{4} \underline{I} 
- \frac{h^2}{8} \underline{B}_{z_{l+2}}(z_{l+1}) \bigg]^{-1},
\end{align}
\begin{align}
\label{eq_curL1llp2}
{\underline{\mathcal{L}}}^{(1)}_{l,l+2}=
& -7 \underline{I}+
2h^2 \underline{V}_{\text{eff}}(z_l)+ \nonumber \\
& \underline{\mathcal{T}}(z_l,z_{l+2}) \underline{s}_{l,l+2} 
\underline{\mathcal{T}}^{T} (z_l,z_{l+2}),
\end{align}
\begin{align}
\label{eq_curL2llp2}
\underline{{\mathcal{L}}}^{(2)}_{l,l+2}=   
\underline{\mathcal{T}}(z_l,z_{l+2})   \left(-\underline{I}+ \underline{s}_{l,l+2} \right)^{T},
\end{align}
\begin{align}
\label{eq_curL3llp2}
\underline{{\mathcal{L}}}^{(3)}_{l,l+2}= -\left( \underline{\mathcal{L}}_{l,l+2}^{(2)} \right) ^T ,
\end{align}
\begin{align}
\label{eq_curL4llp2}
\underline{{\mathcal{L}}}^{(4)}_{l,l+2}= 14 \underline{I}-4h^2 \underline{V}_{\text{eff}}(z_{l+2}) -
\underline{s}_{l,l+2},
\end{align}
\begin{align}
  \label{eq_curly_y}
\underline{\mathcal{Y}}_{0,l+2}=
\left[   \underline{{\mathcal{L}}}^{(4)}_{0,l}  -   
\underline{\mathcal{T}}(z_l,z_{l+2}) \underline{s}_{l,l+2}  
\underline{\mathcal{T}}^T(z_l,z_{l+2})       \right]^{-1},
\end{align}
\begin{align}
  \label{eq_curly_lplus2_1}
\underline{{\mathcal{L}}}^{(1)}_{0,l+2}=  \underline{{\mathcal{L}}}^{(1)}_{0,l} + 
\underline{{\mathcal{L}}}^{(2)}_{0,l}  \underline{\mathcal{Y}}_{0,l+2} 
\left( \underline{{\mathcal{L}}}^{(2)}_{0,l} \right)^T,
\end{align}
\begin{align}
\underline{{\mathcal{L}}}^{(2)}_{0,l+2}= 
  \underline{{\mathcal{L}}}^{(2)}_{0,l}  
\underline{\mathcal{Y}}_{0,l+2}  
\underline{{\mathcal{L}}}^{(2)}_{l,l+2}  ,
\end{align}
\begin{align}
\underline{{\mathcal{L}}}^{(3)}_{0,l+2}=   -
\left( \underline{{\mathcal{L}}}^{(2)}_{l,l+2} \right)^T  
\underline{\mathcal{Y}}_{0,l+2} 
\left( \underline{{\mathcal{L}}}^{(2)}_{0,l} \right)^T ,
\end{align}
\begin{align}
    \label{eq_curly_lplus2_4}
\underline{{\mathcal{L}}}^{(4)}_{0,l+2}=   
\underline{{\mathcal{L}}}^{(4)}_{l,l+2}  - 
\left( \underline{{\mathcal{L}}}^{(2)}_{l,l+2} \right)^T  
\underline{\mathcal{Y}}_{0,l+2} \underline{{\mathcal{L}}}^{(2)}_{l,l+2},
\end{align}
\begin{align}
  \label{eq_normal_l1}
\underline{L}^{(1)}(z_0,z_{l+2})= -\frac{1}{2}\underline{A}+\frac{1}{6h} 
\underline{\mathcal{L}}_{0,l+2}^{(1)},
\end{align}
\begin{align}
\underline{L}^{(2)}(z_0,z_{l+2})= 
\frac{1}{6h} 
\underline{\mathcal{L}}_{0,l+2}^{(2)},
\end{align}
\begin{align}
\underline{L}^{(3)}(z_0,z_{l+2})= 
\frac{1}{6h} 
\underline{\mathcal{L}}_{0,l+2}^{(3)},
\end{align}
and
\begin{align}
    \label{eq_normal_l4}
\underline{L}^{(4)}(z_0,z_{l+2})= -\frac{1}{2}\underline{A}+\frac{1}{6h} 
\underline{\mathcal{L}}_{0,l+2}^{(4)} - 7 \underline{I} + 2 h^2 \underline{V}_{\text{eff}}(z_{l+2}).
\end{align}
Note that the quantity $Y_{0,l+2}$ defined in Ref.~\cite{SymmGLD} 
contains a typo in the non-labeled
equation after Eq.~(71): the 11-element should be $-1$ and not  $1$ and
the 22-element should be $1$ and not $-1$.
If the typo was not corrected, the
plus sign on the right hand side of Eq.~(\ref{eq_curly_lplus2_1}) would
be a minus sign,
and the
minus sign on the right hand side of Eq.~(\ref{eq_curly_lplus2_4}) would
be a plus sign.

With the above definitions, the algorithm reads:
\begin{itemize}
\item Initialization:
\begin{enumerate}
\item Initialize $\underline{\varphi}(z_0)$
and $\underline{\varphi}'(z_0)$ [see Eqs.~(\ref{eq_bc1})-(\ref{eq_bc6})]. 
\item Initialize $\underline{\mathcal{L}}^{(j)}_{0,2}$
($j=1-4$) using Eqs.~(\ref{eq_curL1llp2})-(\ref{eq_curL4llp2}) with $l=0$.
\item If desired, calculate $\underline{L}^{(j)}(z_0,z_2)$ using
  Eqs.~(\ref{eq_normal_l1})-(\ref{eq_normal_l4})
with $l=0$ and then calculate $\underline{\varphi}(z_2)$
and $\underline{\varphi}'(z_2)$ using Eq.~(\ref{eq_OmegarelationB2}) with
$z'=z_0$ and $z''=z_2$. 
\end{enumerate}
\item For $l=2,4,\cdots,2N-2$:
\begin{enumerate}
\item Calculate $\underline{\mathcal{L}}^{(j)}_{l,l+2}$ using
  Eqs.~(\ref{eq_curL1llp2})-(\ref{eq_curL4llp2}).
\item Calculate $\underline{\mathcal{Y}}_{0,l+2}$ using
  Eq.~(\ref{eq_curly_y}).
\item Calculate $\underline{\mathcal{L}}^{(j)}_{0,l+2}$ using
  Eqs.~(\ref{eq_curly_lplus2_1})-(\ref{eq_curly_lplus2_4}).
\item If desired, calculate $\underline{L}^{(j)}(z_0,z_{l+2})$
  using Eqs.~(\ref{eq_normal_l1})-(\ref{eq_normal_l4})
and then calculate $\underline{\varphi}(z_{l+2})$
and $\underline{\varphi}'(z_{l+2})$ using Eq.~(\ref{eq_OmegarelationB2})
with $z'=z_0$ and $z''=z_{l+2}$. 
\end{enumerate}
\end{itemize}

\section{Explicit expressions for $\vec{a}^{(j)}$}
\label{appendix_asymptotics}
The explicit expressions for $\vec{a}^{(j)}(k_{n_{\rho}}^{(j)})$,
introduced in Eqs.~(\ref{eq_fnrhoj}) and (\ref{eq_gnrhoj}),
for $\tilde{\delta}=0$ read
\begin{align}
  \vec{a}^{(1)}(k_{n_{\rho}}^{(1)}) =
\left[ E_{\Omega}^2 + 4 \left| b(k_{n_{\rho}}^{(1)}) \right|^2 \right]^{-1/2}
  \left(
  \begin{array}{c}
    -E_\Omega \\
    \sqrt{2} b(k_{n_{\rho}}^{(1)}) \\
    \sqrt{2} b(k_{n_{\rho}}^{(1)}) \\
    0
    \end{array}
\right),
\end{align}
\begin{align}
  \vec{a}^{(2)}(k_{n_{\rho}}^{(2)}) =
    \frac{1}{\sqrt{2}}
  \left(
  \begin{array}{c}
    0\\
    -1\\
    1\\
    0
    \end{array}
\right),
\end{align}
and
\begin{align}
\label{eq_veca34}
  \vec{a}^{(3/4)}(k_{n_{\rho}}^{(3/4)}) = \nonumber \\
\left[ N ( k_{n_{\rho}}^{(3/4)} ) \right]^{-1/2}
  \left(
  \begin{array}{c}
    \pm 2 b(k_{n_{\rho}}^{(3/4)}) \\
    \pm E_{\Omega}/\sqrt{2}\\
    \pm E_{\Omega}/\sqrt{2}\\
    \sqrt{E_{\Omega}^2+ 4 \left( b(k_{n_{\rho}}^{(3/4)}) \right)^2}
    \end{array}
\right),
\end{align}
where 
\begin{align}
b(k_{n_{\rho}}^{(j)}) = \frac{\hbar^2 k_{\text{so}} k_{n_{\rho}}^{(j)}}{m}
\end{align}
and
\begin{align}
N ( k_{n_{\rho}}^{(3/4)} )= 
E_{\Omega}^2+ 4 \left( b(k_{n_{\rho}}^{(3/4)}) \right)^2+ 
\left| E_{\Omega}^2+ 4 \left( b(k_{n_{\rho}}^{(3/4)}) \right)^2 \right| .
\end{align}
In Eq.~(\ref{eq_veca34}), 
the ``$+$'' and ``$-$'' of the ``$\pm$'' are associated with 
the superscripts 3 
and 4, respectively.

\section{Details related to the ``rotation approach''}
\label{appendix_rotation}

The operator $\hat{R}$ in the singlet-triplet basis
reads
\begin{align}
  \underline{R}=\left(
  \begin{array}{cccc}
    \underline{R}_0 & \underline{0} & \cdots & \underline{0} \\
    \underline{0} & \underline{R}_0 &  &  \\
    \vdots &  & \ddots &  \\
    \underline{0} &  &  & \underline{R}_0
    \end{array}
  \right) ,
\end{align}
where
\begin{align}
  \underline{R}_0= 
\left(
  \begin{array}{cccc}
    \cos(k_{\text{so}}z) & 0 & 0 & -\imath \sin( k_{\text{so}} z) \\
    0 & 1 & 0 & 0  \\
    0 & 0 & 1 & 0 \\
    -\imath \sin( k_{\text{so}} z) & 0 & 0 & \cos(k_{\text{so}}z)
    \end{array}
  \right) .
\end{align}
Operating with $\underline{R}_0^{\dagger}$ on 
the vector that contains the singlet-triplet basis states,
we obtain the rotated basis states $|R_j \rangle$:
\begin{align}
|R_1\rangle=\cos(k_{\text{so}}z) |S_0\rangle- \imath \sin(k_{\text{so}}z)|T_0\rangle , 
\end{align}
\begin{align}
|R_2\rangle= |T_{+1}\rangle , 
\end{align}
\begin{align}
|R_3\rangle= |T_{-1}\rangle , 
\end{align}
and
\begin{align}
|R_4\rangle=-\imath \sin(k_{\text{so}}z) |S_0\rangle+\cos(k_{\text{so}}z) |T_0\rangle
.
\end{align}
In the $|R_j \rangle$ basis,
the operator $\hat{\Sigma}_z^2$ is diagonal with diagonal elements $1$, $0$,
$0$ and $0$. 
This result is used in interpreting the approximate identity
given in Eq.~(\ref{eq_capc_approx}).

Using the basis 
$\{|R_{1}\rangle,\cdots,|R_4 \rangle \}$, the matrix representation of $U$ for 
$\tilde{\delta}=0$ reads
\begin{align}
  \underline{U}=\left(
  \begin{array}{cccc}
    \underline{U}_0 & \underline{0} & \cdots & \underline{0} \\
    \underline{0} & \underline{U}_0 &  &  \\
    \vdots &  & \ddots &  \\
    \underline{0} &  &  & \underline{U}_0
    \end{array}
  \right) ,
\end{align}
where
\begin{align}
\underline{U}_0=\frac{1}{\sqrt{2}}
\begin{pmatrix}
    0  & \sqrt{2} &   0 &  0 \\
-1 & 0 & -\sqrt{2}E_{\Omega} c_+ & \sqrt{2}E_{\Omega} c_- \\
1 & 0 & -\sqrt{2}E_{\Omega} c_+ & \sqrt{2}E_{\Omega} c_- \\
0 & 0 & (E_{\text{so}} + \sqrt{c})c_+ & (-E_{\text{so}} + \sqrt{c})c_- 
\end{pmatrix}
\end{align}
with
\begin{align}
c = (E_{\text{so}})^2 + (2 E_{\Omega})^2
\end{align}
and
\begin{align}
c_{\pm} = \left( c \pm E_{\text{so}} \sqrt{c} \right) ^{-1/2}.
\end{align}
The basis states $|D_j \rangle$, obtained by acting with
$\underline{U}_0^{\dagger}$ on the 
basis states $|R_j \rangle$, read
\begin{align}
|D_1\rangle =  \frac{-1}{\sqrt{2}} \left(|R_2\rangle-|R_3\rangle \right) ,
\end{align}
\begin{align}
|D_2\rangle= |R_1\rangle,
\end{align}
\begin{align}
|D_3\rangle= 
{{c}_{+}} 
\left[-E_{\Omega} \left( |R_2\rangle + |R_3\rangle \right)  + 
\frac{E_{\text{so}}+\sqrt{c}}{\sqrt{2}} |R_4\rangle \right] ,
\end{align}
and
\begin{align}
|D_4\rangle=  {{c}_{-}} \left[E_{\Omega} \left(|R_2\rangle+ |R_3\rangle \right)  + 
\frac{-E_{\text{so}}+\sqrt{c}}{\sqrt{2}}|R_4\rangle \right].
\end{align}

\bibliography{paper_v14}

\begin{thebibliography}{57}
\bibitem[{foo()}]{footnote1}
\bibinfo{note}{To obtain the even-$z$ bound state energies, we use the inline
  equation that is 11 lines below Eq.~(10) of Ref.~\cite{Olshanii2}. To obtain
  the odd-$z$ bound state energies, we use Eq.~(10) of Ref.~\cite{Cui_Ebodd}.
  The bound state energies are then determined self-consistently using the
  energy-dependent scattering length and scattering volume for the finite-range
  Gaussian potential as input. To calculate scattering lengths and scattering
  volumes at negative energies, the Pad{\'e} approximation \cite{Pade} is
  used.}
\end{thebibliography}

\end{document}